\documentclass[aps,amsmath,amssymb,twocolumn]{revtex4}

\usepackage{graphicx}
\usepackage{epstopdf}
\usepackage{hyperref}
\usepackage{epsfig}
\usepackage{color}
\usepackage{psfrag}
\usepackage{float}
\usepackage[squaren,Gray]{SIunits}
\usepackage{bbold}

\begin{document}

\title{A macroscopic object passively cooled into its quantum ground state of motion: 
 beyond single-mode cooling}
\author{D. Cattiaux$^{*}$, I. Golokolenov$^{*}$, S. Kumar$^{*}$, M. Sillanp\"a\"a$^{**}$, L. Mercier de L\'epinay$^{**}$, R. R. Gazizulin$^{*}$, X. Zhou$^{***}$, A. D. Armour$^{****}$, O. Bourgeois$^{*}$, A. Fefferman$^{*}$ and E. Collin$^{*,\dag}$}

\address{(*) Univ. Grenoble Alpes, Institut N\'eel - CNRS UPR2940,  \\
25 rue des Martyrs, BP 166, 38042 Grenoble Cedex 9, France \\
          (**) Departement of Applied Physics, Aalto University,  FI-00076 Aalto, Finland   \\
					          (***) IEMN, Univ. Lille - CNRS UMR8520, 
Av. Henri Poincar\'e, Villeneuve d'Ascq 59650, France \\
 	(****) Centre for the Mathematics and Theoretical Physics of Quantum Non-Equilibrium Systems \\ and School of Physics and Astronomy, University of Nottingham, Nottingham NG7 2RD, United Kingdom 
 }

\date{\today}

\begin{abstract}
$^{\dag}$ Corresponding Author:  {\it eddy.collin$@$neel.cnrs.fr}.
\end{abstract}

\newcommand{\Teff}{D} 

\maketitle

\section{ABSTRACT}

{\bf 
The building blocks of Nature, namely atoms and elementary particles, are described by quantum mechanics. This fundamental theory 
is the ground on which physicists have built their major mathematical models \cite{RefFeynman}. 
Today, the unique features of quantum objects have led to the advent of promising 
quantum technologies \cite{Qubits,Qualgo}.
However, the macroscopic world is 
manifestly
classical, and the nature of the quantum-to-classical crossover remains one of the most challenging open question of Science to date. 
In this respect, moving objects play a specific role \cite{Legett1,Leggett2}.
Pioneering experiments over the last few years have begun exploring quantum behaviour of micron-sized mechanical systems, 
either by passively cooling single GHz modes, or by
adapting laser cooling techniques developed in atomic physics to cool specific modes far below the temperature of their surroundings \cite{Teufel,Painter,Kippen,Aspel,SillanpaaEnt,Groblacher}. 
Here instead we describe a very different approach, {\it passive cooling} of a 
whole %
micromechanical system down to 500$~\mu$K, reducing the average number of quanta in the fundamental vibrational mode 
at 15$~$MHz 
to just 0.3 
(with even lower values expected for higher harmonics); the challenge being to be still able to detect the motion without disturbing the system noticeably. 
With such an approach 
higher harmonics and the surrounding environment are also cooled, leading to potentially much longer mechanical coherence times, 
and enabling 
experiments questioning 
mechanical wave-function collapse \cite{BassiCollapse}, potentially from the gravitational background \cite{BouwmeesterPRL,Penrose}, and
quantum thermodynamics \cite{RefQthermo}.
Beyond the average behaviour, here we also report on the {\it fluctuations} of the fundamental vibrational mode of the device in-equilibrium with the cryostat. 
These reveal a surprisingly complex interplay with the local environment and allow characteristics of two distinct {\it thermodynamic baths} to be probed.

$ $
}

\section{INTRODUCTION}

Centre-of-mass motion stands out in quantum mechanics \cite{Legett1,Leggett2}. It plays a central role in quantum models of gravity \cite{Penrose,Stamp} and is at the core of
 continuous spontaneous localisation (or collapse) models (CSL) \cite{BassiCollapse}. Understanding the quantum behaviour of macroscopic moving objects could thus be the key to unification of quantum mechanics and general relativity, solving at the same time the long-standing issue of the wave-packet reduction postulate interpretation \cite{BookJoos}.
In practice CSL models 
are most effectively
challenged by probing lengthscales of order $10 - 100 ~$nm: within the {\it mesoscopic range} \cite{BassiCollapse}.
Besides, motion is at the core of the basic definition of heat: {\it phonons} arise in quantum mechanics as the quasi-particle describing how atoms move.
It is therefore important to explore practical ways in which thermodynamics could be probed at the quantum level beyond electromagnetic degrees of freedom
\cite{JukkaQuantumPhotons,JukkaQuantumThermo}.

Very few setups can detect motion near the Heisenberg limit, that is with the minimum back-action allowed by quantum mechanics \cite{ClerkPRA, SchwabSET, VanDerZant}. In practice, this can be conveniently realized by coupling the mechanical degree of freedom to an optical cavity \cite{BraginskyManukin, LKB}. The tremendous capabilities of opto-mechanics for force sensing have led recently to gravitational wave detection \cite{LIGO}.
One extremely promising technology is the microwave version of opto-mechanics, where the mechanical element modulates the resonance frequency of an $RLC$ circuit \cite{RegalNatPhys2008,Teufel}. It inherits the properties of conventional opto-mechanics, is directly compatible with quantum electronics technologies \cite{ClerkDevoretSchoelkopf, SillanpaaNEMSQubit}, with the great advantage of low energy photons being more compatible with cryogenic setups \cite{Teufel, SillanpaaEnt,PRXSchwab}.

Conventionally,
a mechanical mode can be said in its quantum ground state when its thermal population $n(T) < 1$, which can be achieved by lowering the temperature $T$. 
Alternatively, the high degree of control reached within optomechanical systems 
enables the use of back-action cooling \cite{Braginsky,Teufel}. 
This comes at a cost: the mode is strongly damped by the light field, while the surrounding bath remains warm \cite{Aspel,steele}. 
GHz acoustics has been passively cooled to the ground state using conventional dilution cryostats \cite{Oconnell,SAW,PainterPhonoCryst}.
However, these systems are limited to extremely small centre-of-mass displacements ({\it zero} in the case of breathing modes \cite{Oconnell,PainterPhonoCryst}), 
even though they do contain a very large number of individual atoms. 
 Hence, 
they are not suitable for tests of quantum decoherence and collapse theories \cite{Legett1}.

Mechanical modes within micro/nano-mechanical systems with macrocopic masses and which can tolerate large motional amplitudes have resonance frequencies in the MHz range. 
Their passive cooling 
therefore necessitates 
sub-milliKelvin temperatures, which is the range attained by ultimate cryogenics: {\it nuclear adiabatic demagnetisation}. The ability to measure the system without disturbing it, and the demonstration of its thermal equilibrium with the cryostat are both significant challenges \cite{pickett}.

Here we report on in-equilibrium ground-state cooling of a whole 15-micron diameter aluminium 
mechanical device resonating at $\omega_m=2 \pi \times 15.1~$MHz in its first flexure and coupled to a $\omega_c=2 \pi \times 5.7~$GHz microwave cavity, installed on a nuclear demagnetization cryostat reaching about 500$~\mu$K. Remarkably, no signs of thermal saturation are detected in the mechanical properties as shall be discussed below. 
The mechanical device consists of a 50$~$nm vacuum-gap capacitor (see Fig. \ref{fig_1}) which forms a drum-head \cite{SillanpaaEnt}, and the cryogenic setup consists of a home-made laminar copper nuclear stage mounted on a wet dilution cryostat (see Ref. \cite{ZhouPRAppl} for details).
The single-phonon opto-mechanical coupling $g_0$ for these modes is measured to be about $2 \pi \times 230~$Hz. Single-tone microwave opto-mechanics is used here only for motion detection \cite{AKMreview}. A pump tone 
(of power $P_{in}$) is applied at an angular frequency $\omega_c+\omega_m $ (blue-detuned scheme), or conversely at $\omega_c-\omega_m$  (red-detuned scheme).
The mechanical motion generates a sideband peak in the output spectrum at the cavity frequency $\omega_c$. 
Energy is transferred between the mechanical mode and the microwave field, resulting in either enhanced damping of the motion as we increase the drive power (red-detuning), or  amplification (blue-detuning).
Measurements are performed at the lowest possible powers 
($n_{cav} \propto P_{in}$ about 300 - 600 drive photons confined in the cavity) in order to limit the impact of damping/amplifying. 
However, driving the system up to the blue-detuned instability (when the damping vanishes), we can show that the mechanical mode exhibits self-sustained motion in the nanometer range \cite{CiteCattiauxPRR}, which 
means that these devices are potentially very well adapted for exploring CSL physics.
Furthermore, {\it no physical heating} of the device can be detected in this range of injected powers, down to $500~\mu$K, which is remarkable. Optical systems are usually limited by their ability to feed-in energy from the photon field, making pulsed experiments mandatory \cite{PainterPhonoCryst,davisTLS}.
 Details on the setup and measurements can be found in Supplementary Information.

\section{RESULTS}

\begin{figure}
			 \includegraphics[width=12.cm]{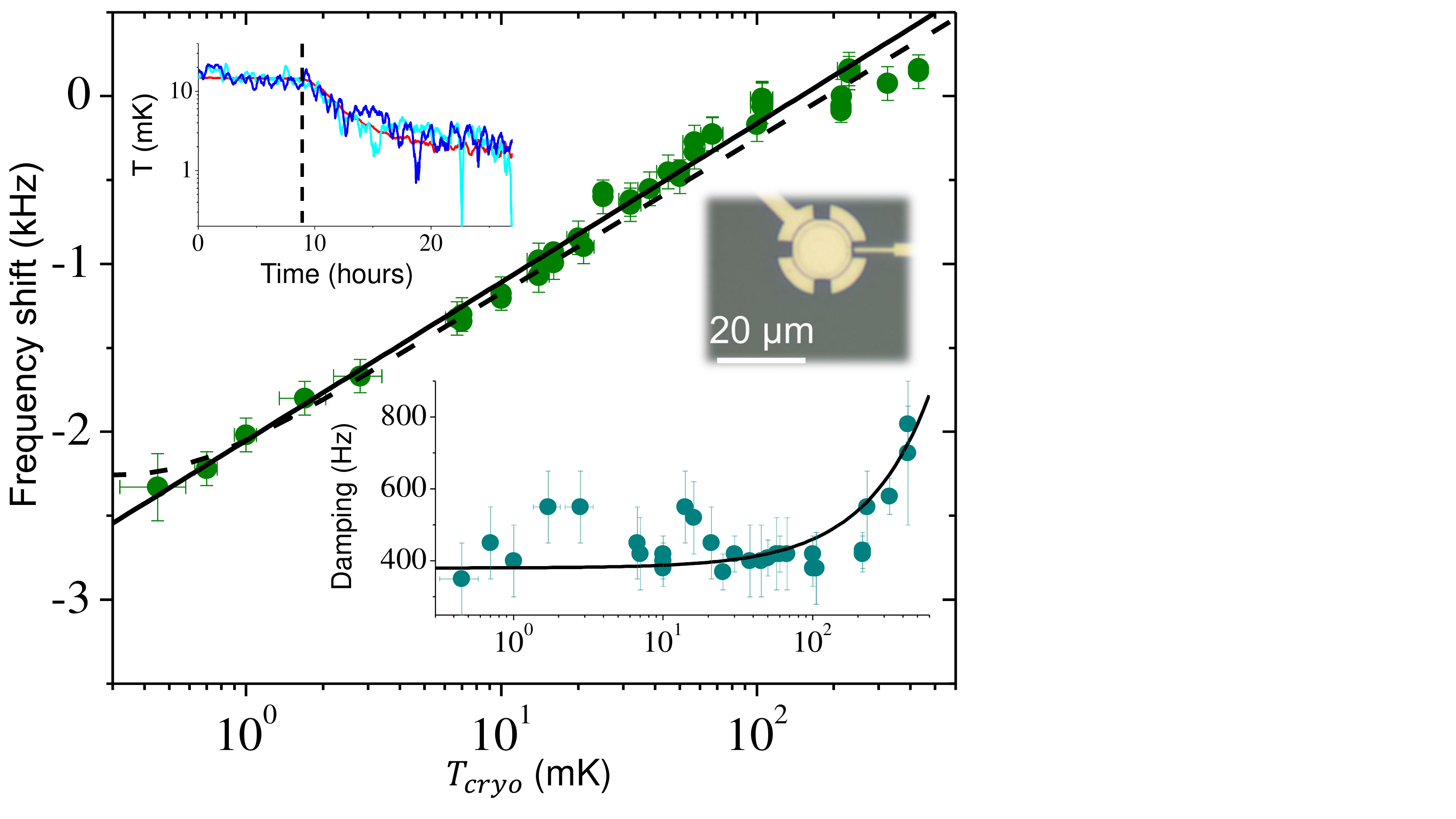}			
\vspace*{-5 mm}
			\caption{\small{{\bf Mechanical properties versus temperature.} Main graph: mechanical mode frequency shift $(\omega_m [T]-\omega_0)/(2 \pi)$ [with respect to $\omega_0/(2 \pi) \approx 15.1~$MHz] as a function of cryostat temperature. The black line is a simple Logarithmic fit, while the dashed line is the full theoretical fit function (see text). An optical picture of a device from the same batch is also shown. Bottom inset: corresponding mechanical damping $\Gamma_m/(2 \pi)$. The black line is a linear fit (see text). Top inset: measured temperatures at beginning of cooling in a demagnetization process (starting at the dashed vertical). The red curve corresponds to the cryostat $T_{cryo}$, while the two blue curves are mechanical mode temperatures $T_{mode}$ calculated from the spectrum (from two distinct runs, see text).}}
			\label{fig_1}
		\end{figure}			

Formally, three distinct temperatures have to be considered here: the cryostat $T_{cryo}$, the fundamental mechanical mode $T_{mode}$, and the baths $T_{baths}$ directly in contact with it \cite{ZhouPRAppl}. 
Demonstrating that thermal equilibrium is maintained across these different subsystems is the key challenge for passive cooling, which we shall address first. Furthermore, characterizing the properties of the baths that couple to the mechanical mode is of itself a significant issue. 
This is our second topic, demonstrating the capabilities of our microwave/microkelvin platform. \\

In Fig. \ref{fig_1} we present the mechanical frequency shift $\omega_m (T)-\omega_0$ (with respect to the high-temperature value $\omega_0$, main graph) and the damping $\Gamma_m$ (bottom inset) of the first flexural mode, as a function of cryostat temperature $T_{cryo}$.
These parameters are obtained from Lorentzian fits of the measured sideband peak (position and linewidth, respectively).
Below 100$~$mK, the mechanical damping saturates which is a signature of clamping losses \cite{RefClampLoss,RefClampLoss2}. 
This means that ``phonon tunneling'' dominates energy relaxation, and therefore one of the baths to be considered is {\it the phonons} surrounding the drum device.
On the other hand, the frequency shift follows a Logarithmic temperature dependence in the whole range, as is commonly observed with low temperature nanomechanics \cite{cite_Mohanty}. 
This is interpreted as the signature of Two-Level Systems (TLS) defects present in the structure, 
that couple to the macroscopic mode via the strain field \cite{RefGlassULT2,cite_Kunal}. The exact fit expression from theory is shown as the dashed line in Fig. \ref{fig_1} (see Supplementary Information for details) \cite{ReviewPhilips}. This leads us to identify a second bath coupled to the mechanics, namely {\it the TLSs}.
The absence of thermal saturation on $\omega_m (T)$ proves that this bath does cool to the lowest temperatures \cite{ZhouPRAppl}.
Since it thermalises through the bulk phonons themselves \cite{ReviewPhilips}, the surrounding phonon bath has to be cold as well. We can thus infer $T_{TLS} \approx  T_{phonons}\approx T_{baths} \simeq T_{cryo}$.

		\begin{figure*}[t!]		 
			 \includegraphics[scale=0.6]{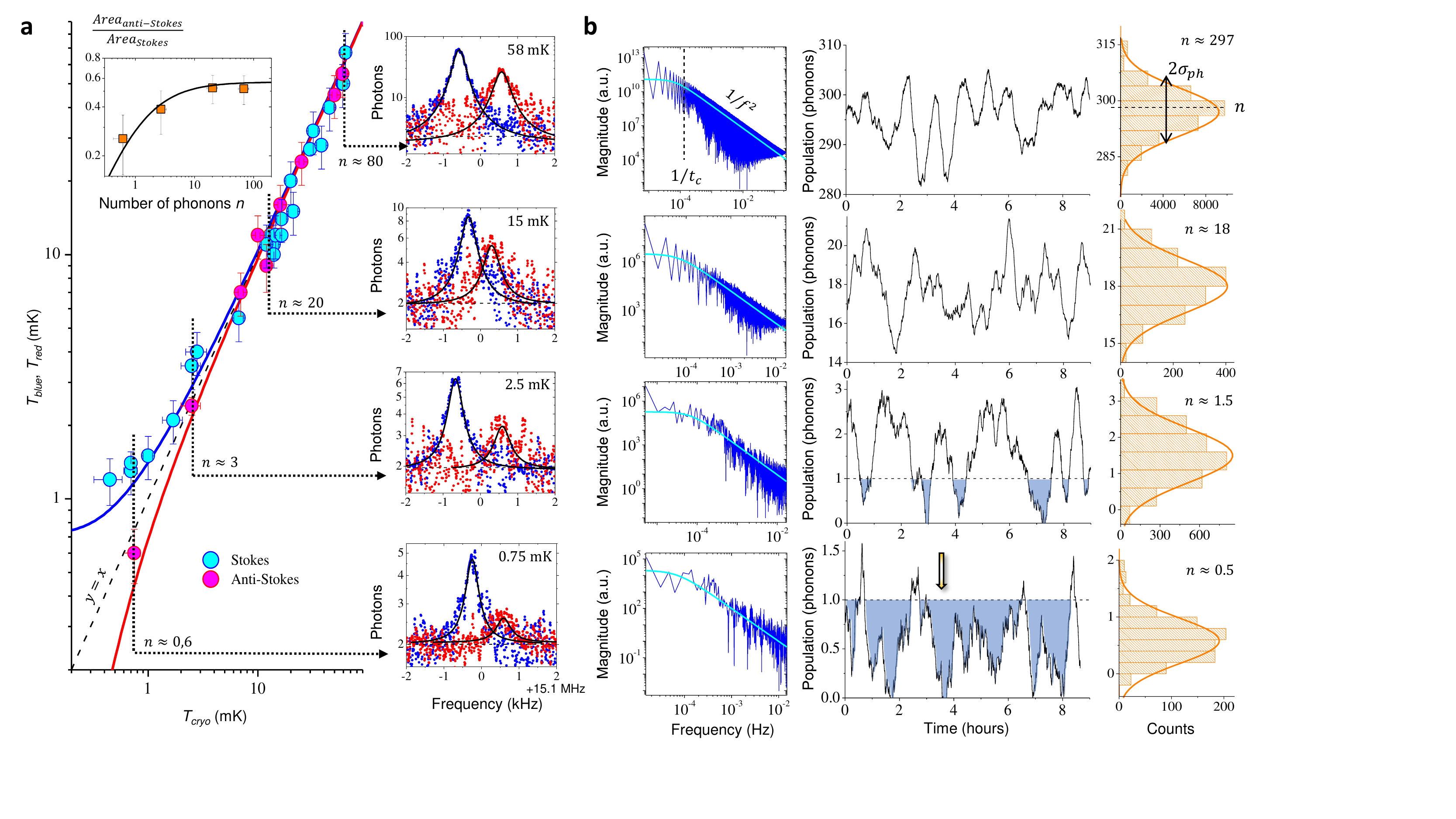}	
			\vspace*{-10 mm}
			\caption{\small{ {\bf Passive ground state cooling.} a, Main graph: the peak area of blue and red sideband pumping measurements leads to two different $T_{blue},T_{red}$  estimates, $\propto n+1$ and $\propto n$ (blue and red lines) respectively, a phenomenon known as sideband asymmetry \cite{PRXSchwab}. Right insets: sideband peaks (blue and red colour for Stokes and anti-Stokes) measured for four different temperatures, with their Lorentzian fits (vertically shifted to 2-photon baseline for readability). The slight shift in position 
(with respect to the actual $\omega_m[T]$) 
is due to a small remnant optical spring due to imperfect tuning of the pump frequency, while the difference in peak heights at high temperature is due to the small amplifying/de-amplifying occurring at the finite input power we use (see text and Supplementary Information). The smallest Anti-Stokes peak required 4 days of continuous averaging. Left top inset: ratio of anti-Stokes and Stokes peak areas enabling sideband asymmetry thermometry (the black line is the theoretical expectation). b, Measurement of in-equilibrium population fluctuations at 220, 13, 1.4 and 0.65$~$mK (top to bottom) obtained with 20 minutes averaging time per point. Middle: time traces (9 h span) demonstrating very slow (and large, see also inset Fig. 1) fluctuations. The shaded areas delimit the regions below 1 phonon. Left: spectrum (FFT transform of autocorrelation) showing a $1/f^2$ type dependence (full line fit) with a low-frequency cutoff $1/t_c$ (dashed vertical). Right: corresponding histograms from which average $n$ and standard deviation $\sigma_{ph}$ can be defined (with Gaussian fit displayed; the slight negativity for the two lowest graphs is due to the finite precision of the analysis procedure, see text and Supplementary Information). 
Error bars mostly arise from the finite stability of mechanical parameters (frequency, damping, see Fig. \ref{fig_1}) and are explained in Supplementary Information.
Note the arrow on the central lowest time-trace that points to a time slot where {\it the measured population drops below our resolution} for about 5 minutes (see text).}}
			\label{fig_2}
		\end{figure*}	

 Cooling from about 10$~$mK to the lowest temperature takes about 2 to 3 days; the beginning of the process is shown in Fig. \ref{fig_1} top inset. 
The mechanical mode temperature $T_{mode}$ is inferred from the area of the measured sideband peak 
using a blue-detuned pumping, as explained below. 
We clearly see that, {\it on average}, the drum-head mode and the cryostat temperatures follow each other, in good equilibrium. 
Furthermore, we can also see that the mechanics reproducibly displays both {\it very large, and very
slow} temperature fluctuations. In the following we will start by discussing the averaged behaviour in detail before going on to describe the properties of the fluctuations.

The sideband signal 
encodes
the position-fluctuations spectrum of the mechanical mode.
At low drive powers, the area of the measured peaks with blue and red pumping (Stokes and Anti-Stokes sidebands) are proportional to $P_{in} \times (n+1)$ and $P_{in} \times(n)$ respectively, with $n(T_{mode})=1/(Exp[(\hbar \omega_m)/(k_B T_{mode})]-1)$ the Bose-Einstein population \cite{AKMreview, PRXSchwab}. 
It is therefore convenient for the experimentalist to define 
from these areas 
two ``effective mode temperatures'':

$T_{blue}=(n+1) (\hbar \omega_m)/k_B $ for blue-detuned pumping, measuring Stokes peak,

$T_{red}= (n) (\hbar \omega_m)/k_B$ for red-detuned pumping, measuring anti-Stokes peak.

We will in the following demonstrate thermalisation of $T_{mode}$ by studying $T_{blue}, T_{red}$ and their {\it ratio} as a function of $T_{cryo}$.
In Fig. \ref{fig_2} a (right insets) we show 
sideband spectra measured at different cryostat temperatures and the same drive power (about 300 photons), with their Lorentzian fits. For this microwave input, a slight amplifying/de-amplifying exists that is visible from the peak height difference of the high temperature spectra (blue and red colour for Stokes and anti-Stokes). In order to extract the genuine small-drive limit, we measure the power dependence for each temperature and pumping scheme. This enables us to plot in Fig. \ref{fig_2} a (main graph) the $T_{blue}, T_{red}$ dependencies with respect to $T_{cryo}$. 
Lines are theoretical calculations, with no free parameters.
At high temperature, $T_{blue} \approx T_{red} \approx T_{cryo}$, as it should; 
this is essentially the limit of conventional experiments using commercial dilution cryostats \cite{ZhouPRAppl}, and justifies the wording ``effective temperature''. \\

However below typically 10$~$mK, sideband asymmetry is visible:
for blue pumping, $T_{blue}$ saturates while for red $T_{red}$ vanishes (see fits in Fig. \ref{fig_2} a main graph). 
This effect is a signature of zero-point-fluctuations in the microwave cavity \cite{PRXSchwab}. 
While $T_{blue}, T_{red}$ are 
no longer simply proportional to
$T_{cryo}$, the asymmetry leads to an in-built primary thermometry by plotting the ratio of Anti-Stokes over Stokes peak areas \cite{AssymThermo}. 
Indeed, in the ideal case of low powers this reduces to $n/(n+1)$; for the finite microwave drive amplitude used for the peaks displayed, the function is renormalized by the ratio of the amplified/de-amplified peak linewidths. This is shown in the top left inset, Fig. \ref{fig_2} a; with the black line being theory with no free parameters. Technical details are given in Supplementary Information. 	
The experiment therefore demonstrates cooling down to an average population for the first flexure of 0.3 quanta (c.f. the lowest Stokes data point in Fig. \ref{fig_2} a). 
A discussion on the thermal modeling of the device can be found in Supplementary Information.

Although the long-time average mode temperature always matches that of the cryostat to within measurement uncertainties, we observe strong fluctuations that occur over a surprisingly long timescale, as mentioned earlier.  Landauer famously remarked "the noise is the signal" \cite{landauer}, and in this case the fluctuations provide important information about the complexity of the thermal environment surrounding the mechanics.
 To characterize them, we acquire continuously (at fixed drive power, 600 photons, and $T_{cryo}$) sideband spectra using the blue-detuned pumping scheme at a reasonably high repetition rate (typically 1 second per file), and then post-process these data with a sliding average of window 20 minutes that produces a fitable peak \cite{ZhouPRAppl}. From the fit we can extract peak position, width and area as a function of time. The area can then be converted (from the calibrated power-dependence) into mode temperature (top inset, Fig. \ref{fig_1}), or population (Fig. 2 b centre) \cite{ZhouPRAppl}. Width and position shall be discussed later in this article.
While a true reconstruction of the phonon tunnelling statistics is out of the scope of the present paper (the measurement is {\it not} at the single phonon level), what we obtain is a sort of ``smoothed'' estimator of the mode occupation number. 
Furthermore, since the finite measuring power disturbs the system we must always correct for this to obtain the unpumped behaviour. 
Errors associated with the correction process explain
the slight negativity of the lowest histograms in Fig. \ref{fig_2} b. These issues are discussed explicitly below and in Supplementary Information.

\begin{figure}[t!]		 
			 \includegraphics[width=11cm]{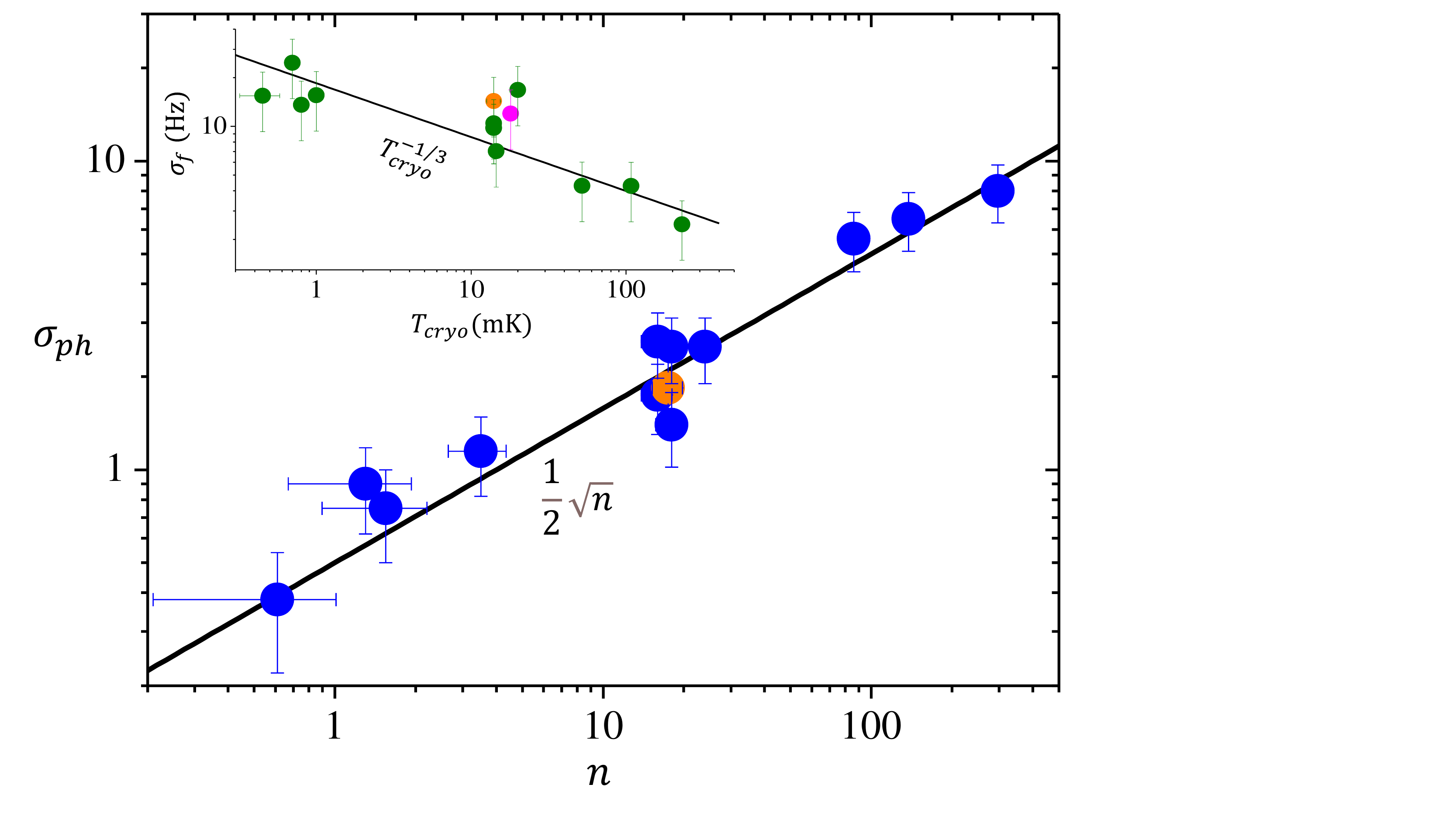}			
			\caption{\small{  {\bf Frequency and population standard deviations.} Main: standard deviation of the phonon population extracted from Fig. \ref{fig_2} b as a function of average population. The black line is a $\propto \sqrt{n}$  
guide to the eye.
Top inset: frequency shift standard deviation calculated over 10 h averaging as a function of cryostat temperature. The line is a power law fit. 
The magenta point is inferred from data taken in the self-oscillating regime; in both graphs the orange dot is obtained with a measuring input power twice smaller (see text). }}
			\label{fig_3}
		\end{figure}	

From time-traces we compute the fluctuation’s spectra (Fig. \ref{fig_2} b left).
These can be fit by a Lorentzian shape: producing a $1/f^2$ dependence ({\it Brown noise}) above a cutoff frequency $1/t_c$ (see full lines). 
Such a spectrum is reminiscent of an {\it Ornstein-Uhlenbeck} process \cite{OUprocess}. The question that arises then is: from what thermodynamic bath?
We expect fluctuations due to coupling to the phonon bath 
to have a characteristic timescale $\tau_{m} = 2/\Gamma_m \approx 1~$ms, the mechanical relaxation time.
This timescale is clearly outside of our bandwidth (which is at best about 1$~$Hz); besides, we observe that the time $t_c \gg \tau_m$ is of the order of $5 (\pm 2)$ hours, and not specifically dependent on temperature (see Fig. \ref{fig_2} b left).
Therefore, our experiment {\it does not} directly 
probe phonon fluctuations.
It is thus natural to consider the second identified bath: {\it TLSs}.
As shall be discussed below, from the frequency (and damping) fluctuations we infer that TLSs are certainly involved, in a complex and specific way. 
And indeed, ultra-low temperature experiments that studied their dynamics report on extremely slow timescales, which are consistent with our $t_c$ \cite{parpiaSpikes}.
A detailed discussion on the statistical treatment is given in Supplementary Information.		

At low temperatures the fluctuations can be especially interesting. For example, note the arrow 
in Fig. \ref{fig_2} b centre that points to a time slot where the measured mode population stays at essentially zero over about 5 minutes. 
Knowing that high frequency modes do cool down to dilution temperatures with no sign of thermal saturation \cite{Oconnell, Painter}, higher frequency modes of the structure have to be empty with a high probability.
 Making an analogy with electronic transport, this result is reminiscent of the one of Ref. \cite{Jukka_NO_QP} that demonstrated the absence of any quasi-particles in a superconductor over seconds.

The statistical distributions computed from the time-traces look reasonably Gaussian (Fig. \ref{fig_2} b right). 
We can then extract the average population $n$ and its standard deviation $\sigma_{ph}$. This is shown in Fig. \ref{fig_3} (main).
Remarkably, 
 we find that the behaviour of $\sigma_{ph}$ is well-described by $0.5\sqrt{n}$, implying a {\it sub-Poissonian} mechanism.
Note that this fit extends from the truly quantum range up to the classical one, without showing any crossover behaviour. 
This level of fluctuations is smaller than what we would expect in terms of the fluctuations in the average energy of a system coupled to a simple thermal bath \cite{thermoADA,RefHO}. However, this is not surprising since fluctuations associated with the phonon bath are effectively averaged over in the measurement. The observed fluctuations are thus most likely a signature of the (potentially rather complex \cite{burnett,TLS_report}) mode-TLS interaction.

The behaviour of $\sigma_{ph}$ is 
completely different from frequency and damping noises measured at the same time. They both display {\it true} $1/f^2$ spectra (with no low frequency cutoff observed), and shall thus be characterized for a given time span (so-called Allan deviation) \cite{MailletACSNano}. Their temperature dependence is similar, and reversed from the one of phonon population, see inset of Fig. \ref{fig_3} presenting the frequency standard deviation $\sigma_{f}$: it {\it grows as we cool}. 
This resembles what has been reported on superconducting cavities, and attributed to TLSs present in the constitutive materials
\cite{burnett}; here, it would be the TLSs generating frequency shifts from strain-coupling \cite{cite_Kunal,RefGlassULT,RefGlassULT2}. 
Details on the statistical analysis can be consulted in Supplementary Information.

Before concluding, we should point out that, as for any measurement, even if the detection power is kept very low it is nonetheless non-zero.
There is therefore a finite {\it back-action} from the detection scheme onto the mechanical mode: the amplifying/de-amplifying seen in the spectra of Fig. \ref{fig_2} a, plus a small cavity noise.
It is rather straightforward to correct for these, and recalculate {\it average} phonon populations \cite{ZhouPRAppl,AKMreview}.  
It is however perfectly natural to wonder how the finite measurement drive affects the {\it fluctuations} themselves, 
in magnitude,  spectral properties and distribution shape. 
In order to verify experimentally that 
recalculated fluctuation characteristics do not depend on pump power, 
we measured a time-trace at half the drive and processed it 
in the same way: this leads to the orange dot in Fig. \ref{fig_3}, which is in agreement with all other data. A thorough discussion on the impact of the measurement scheme can be found in Supplementary Information.\\ 

\section{DISCUSSION}	

In conclusion, we have demonstrated ground-state cooling of a {\it whole} drum-head device
(of mass $5 \times 10^{-14}~$kg),
 an object that can move its centre-of-mass substantially (typically tens of nm in self-sustained motion 
of first flexure),
 from {\it in-equilibrium} properties of the lowest frequency mode. 
 Fluctuations in the occupation of the mechanical mode reveal the complexity of the {\it baths} to which it is coupled.
The microwave/microkelvin platform developed for this experiment opens the path to a unique new regime for experimentalists: quantum thermodynamics can be addressed with phonons through mesoscopic equilibrium properties near the ground state of motion, instead of electrons (or photons) \cite{JukkaQuantumPhotons, JukkaQuantumThermo}.
The stochastic nature of quantum heat transport can be studied from one of the collective, macroscopic mechanical  degrees of freedom of the device towards the continuum formed by the substrate, through the constriction made by the clamping region.
``Conventional'' baths (bulk phonons, Two Level Systems) are thermalized; this is mandatory for thermodynamics, but also a unique advantage for the unraveling of new contributions postulated in CSL models \cite{BouwmeesterPRL,BassiCollapse}. 
This is a feature which is absent for other ground-state cooling platforms 
focused on macroscopic motion 
\cite{Teufel,Aspel}.
Besides, modal-coupling can be controlled with {\it all modes} being cold, down to the lowest frequency one: mechanical decoherence from nearby ``hot modes'' (i.e. with a large thermal population) is thus avoided \cite{OLIVEModeC, DykmanModeC}. This is a unique advantage of having a whole object ground-state cooled, in comparison with a single working mode as is realized in {\it all} other systems \cite{Teufel,Oconnell,SAW,PainterPhonoCryst,Aspel,steele}.

Passive cooling 
allows the ground state to be reached
while preserving the mechanical $Q$. This potentially provides much longer mechanical coherence times, 
enabling unique sensitivities for force sensing to be attained
 \cite{Caves}; the figure of merit being $\frac{\Gamma_m}{2 \pi} \times n \approx 100$ for us at 500$~\micro$K,  
with plenty of scope for improvements 
in $\Gamma_m $ (see e.g. Refs. \cite{Teufel,Schwab7mK}).
The unexpected properties of the fluctuations reported here call for
theoretical input on both thermodynamics and the constitutive matter \cite{RefQthermo,RefGlassULT,RefGlassULT2,parpiaSpikes}.
As for superconducting mesoscopic electronic devices \cite{TLS_report}, Two Level Systems are certainly the key to the understanding of the complex microscopic environment interacting with the mechanical mode.

The present work demonstrates the compatibility between microwave optomechanics and ultra-low temperatures. The next generation of experiments will incorporate a TWPA (Travelling Wave Parametric Amplifier) in order to open the detection bandwidth, potentially down to the phonon relaxation time, while reaching the quantum limit \cite{JPALehnert}. This would enable the detection of single phonon jumps in-and-out of the mechanical mode, similarly to electrons in an SET (Single Electron Transistor), which represents the "holy-grail" of quantum thermal transport experiments.
Besides, the microwave circuitry is fully compatible with standard quantum electronics; which means that future developments will also incorporate a quantum bit \cite{SillanpaaNEMSQubit}. This would enable experiments directly focused on the study of quantum mechanical decoherence, as proposed e.g. in Ref. 
\cite{ArmourBlencoweNJP}. Such proposals that rely on macroscopic motion can only be implemented on low frequency devices \cite{bouwmeesterDM}, as opposed to GHz modes.


\section*{ACKNOWLEDGEMENTS}
We wish to thank O. Maillet, A. Heidmann, P. Verlot and F. Marquardt for very useful discussions.
We acknowledge support from the ERC CoG grant ULT-NEMS No. 647917 (E.C.), StG grant UNIGLASS No. 714692 (A.F.), the STaRS-MOC project from {\it R\'egion Hauts-de-France} and ISITE-MOST project (X.Z.). A.D.A. was supported through a Leverhulme Trust Research Project Grant (RPG-2018-213), and M.S. was supported by the Academy of Finland (contracts 308290, 307757, 312057), by the European Research Council (615755-CAVITYQPD), and by the Aalto Centre for Quantum Engineering. The work was performed as part of the Academy of Finland Centre of Excellence program (project 312057). We acknowledge funding from the European Union's Horizon 2020 research and innovation program under grant agreement No.~732894 (FETPRO HOT).
We acknowledge the use of the N\'eel {\it Cryogenics} facility.
The research leading to these results has received funding from the European Union's Horizon 2020 Research and Innovation Programme, under grant agreement No. 824109, the European Microkelvin Platform (EMP).

\appendix
\section*{AUTHOR CONTRIBUTIONS}
D.C. ran the experiment and made all measurements. M.S. and L. M. de l'E. designed and fabricated the sample. X.Z. installed and calibrated the whole microwave platform, and R.G. took care of the cryogenics. I.G. and S.K. made preliminary microwave experiments. O.B. analysed and modelled the thermal aspects of the experiment. E.C. designed the experiment. A.D.A., A.F. and E.C. supervised the experiment and the data analysis. All authors participated in the writing of the manuscript.

\section*{COMPETING INTERESTS}
The authors declare no competing interests.

\bibliographystyle{ieeetran}

\newpage
$\,$

\newpage
$\,$

\begin{center}

{\huge SUPPLEMENTARY INFORMATION for \\}
{\huge A macroscopic object passively cooled into its quantum ground state of motion: beyond single-mode cooling}
\end{center}

\section{Setup}
\label{setup}

\subsection{Microwave wiring}

\begin{figure}[h!]		 
			 \includegraphics[width=13cm]{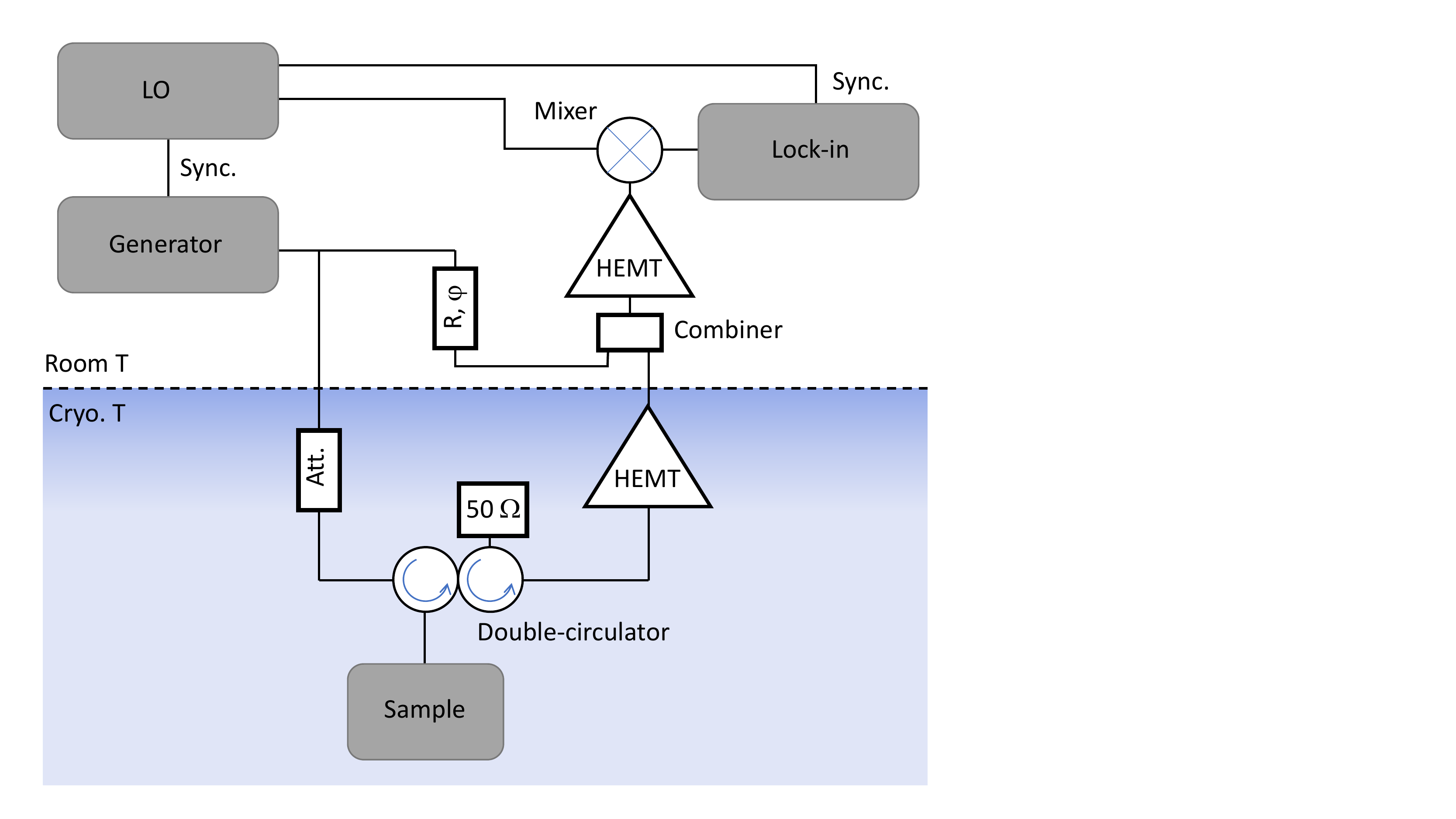}			
			\caption{\small{  {\bf Microwave circuit schematic.} Simplified circuitry of the microwave setup (note attenuation, circulators with 50$~\Omega$ load, $R-\phi$ opposition settings); see Ref. \cite{ZhouPRAppl} for details. }}
			\label{fig_4}
		\end{figure}
		
The microwave setup is a completely conventional one. 
Two platforms have been equipped and calibrated:  a commercial dry Bluefors\textregistered$~$ cryostat for preliminary measurements (not presented in the core of the manuscript), and the home-made wet nuclear demagnetization cryostat.
On the Bluefors\textregistered, a LNF\textregistered$~$ cryo-HEMT is present on the 4$~$K stage with a power combiner mounted in front of it allowing to cancel the strong pump tone (avoiding amplifier saturation). On the nuclear demag. it is a Caltech\textregistered$~$ cryo-HEMT bolted on the 4$~$K stage. Both have a second room temperature LNF\textregistered$~$ HEMT amplifier. On the nuclear demag. cryostat the ``opposition line'' made with the power combiner is at room temperature; the Caltech amplifier is linear enough so that it never saturates with the pump powers we use (from an Agilent\textregistered$~$ 20$~$GHz synthesiser). After mixing up the signal with a local tone (LO) shifted by $\pm \Omega_m + 2~$MHz, the measurement is performed with a Zurich Instrument\textregistered$~$ lock-in that is demodulated at the shift frequency (used in spectrum mode).
A basic drawing is provided in Fig. \ref{fig_4}. 
The coaxial lines from room temperature to 4$~$K are stainless steel, and all the cryogenic routing below is NbTi apart from the mixing chamber level which is copper.
Details with complete schematics can be found in Ref. \cite{ZhouPRAppl} and Ref. \cite{CattiauxThese}. The noise background (brought back at input of the cryogenic HEMT) is on the nuclear demag. setup about 100$~$photons in this experiment. Note that this is clearly not single-shot, but the circuit has the valuable advantage of being simple.

\begin{figure}[h!]		 
			 \includegraphics[width=16cm]{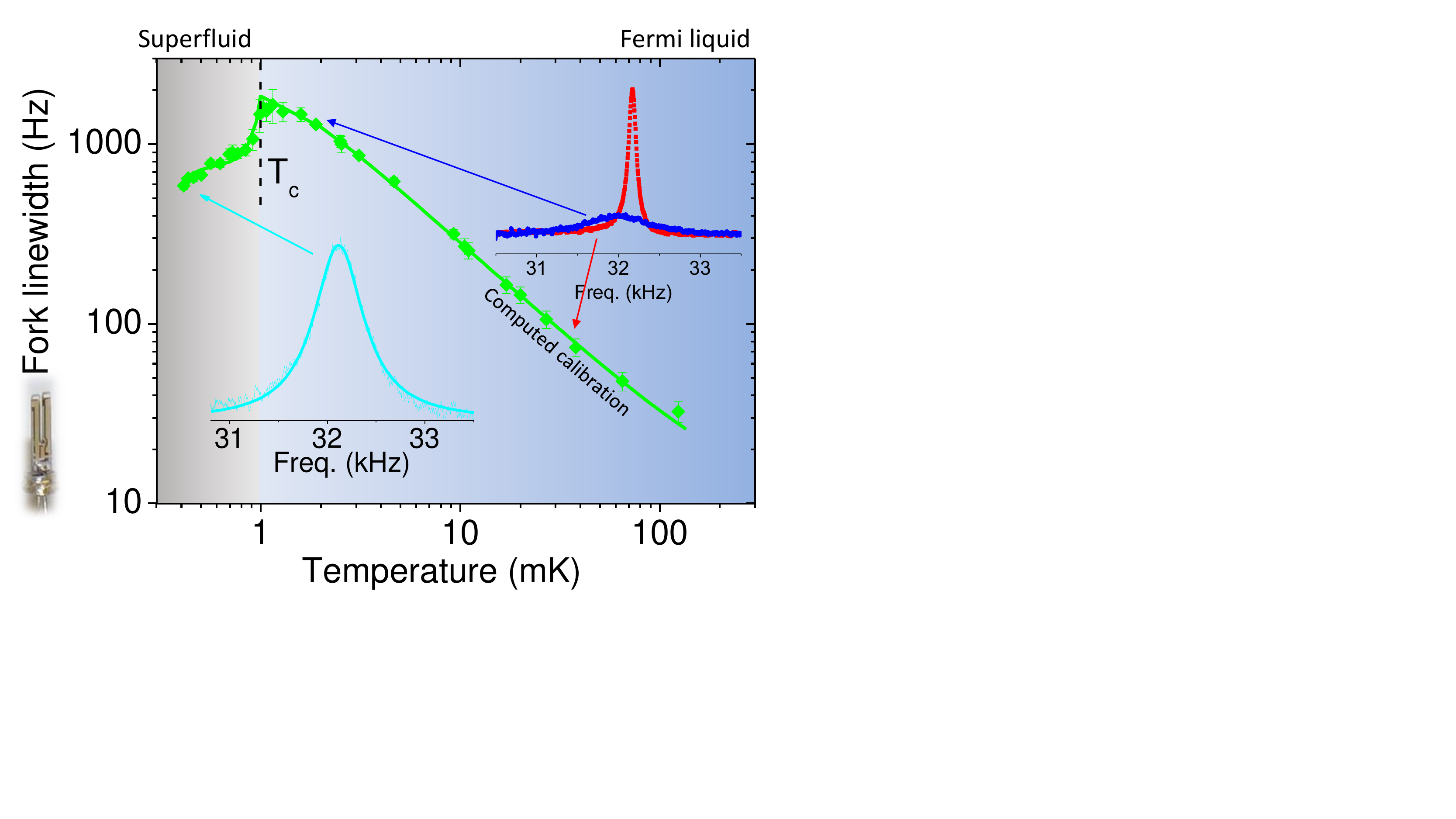}	
					\vspace*{-2cm}
			\caption{\small{  {\bf Tuning fork $^3$He thermometry.} Main: temperature dependence of the damping (in Hz) measured on the tuning fork. Insets: Lorentzian resonance peaks of the tuning fork corresponding to different working points. See Ref. \cite{ZhouPRAppl} for details. }}
			\label{fig_5}
		\end{figure}

\subsection{Cryogenics}

The nuclear demagnetization cryostat is made of two stages: a (conventional) home-made dilution unit that reaches base temperature of 10$~$mK, followed by a copper laminar nuclear stage on which the experimental cell is mounted \cite{paperYuriSatge}. The two stages are connected to each other through a Lancaster-made aluminium heat switch \cite{lancHeatSwtch}. 
On the dilution unit, the temperature is inferred from a calibrated carbon resistor and a commercial magnetic field fluctuation thermometer (MFFT) from Magnicon\textregistered. On the nuclear stage we use a home-made $^3$He thermometer: a quartz tuning fork immersed in the fluid that measures its viscosity \cite{RefFork}. This (almost) primary thermometry also provides a fixed temperature point: the superfluid transition $T_c$ at 0.95$~$mK (for 0$~$bar pressure).  
Calibration of the thermometer is shown in Fig. \ref{fig_5}. 
Note that near $T_c$ the fork is very strongly damped, and the sensitivity is rather poor; however above and below, we can infer temperature very accurately.
Details can be looked at in Ref. \cite{ZhouPRAppl} and Ref. \cite{CattiauxThese}.

The ultra-low temperature cooling process is single shot, and starts with a precooling period of about a week: the heat switch is closed (a small magnetic field is applied on the aluminium heat switch to make it normal), and the copper is cooled to (almost) the mixing chamber temperature while a 7$~$T field is applied to it. 
The coil producing this field is well compensated, and the cell is surrounded by a superconducting lead shield. 
After that, the heat switch is opened (the aluminium is made superconducting by removing the small field in the switch), and the large field of 7$~$T is slowly decreased down to a final value $B_{fin}$ (always larger than 100$~$mT); this makes the nuclear spins of the copper atoms cool down, which in turn cool down the conduction electrons. The cryostat can stay cold for about a week at an almost fixed temperature (see e.g. Ref. \cite{PobellBook} for details on the technique). The beginning of the cooling process is shown in Fig. \ref{fig_1} top inset of the paper.   

\begin{figure}[h!]		
			 \includegraphics[width=24cm]{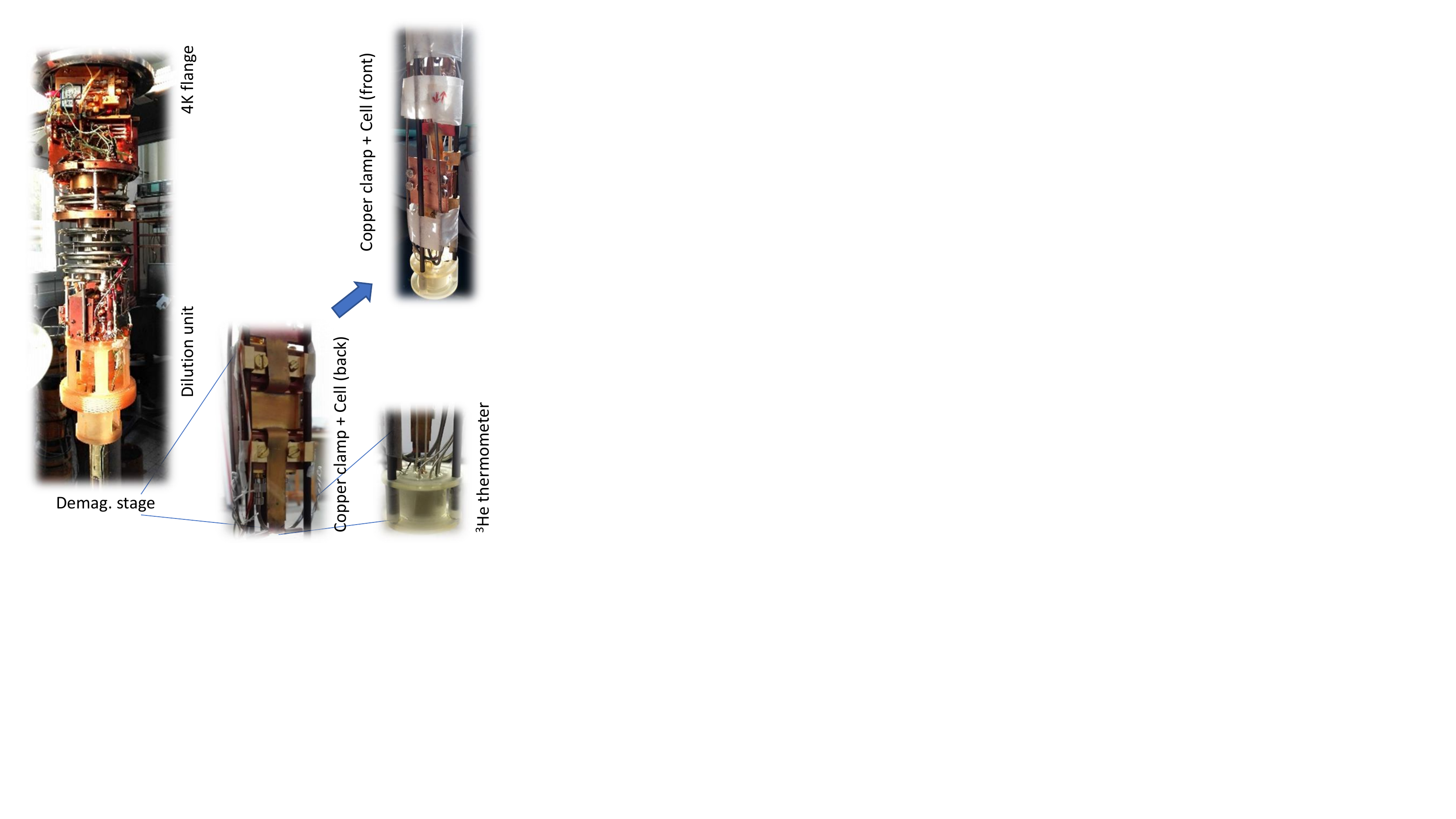}			
\vspace*{-4cm}
			\caption{\small{  {\bf Nuclear demag. cryostat and copper cell.} Left: picture of the dilution unit stage of the nuclear demagnetisation cryostat. Right: bottom of the nuclear stage ({\it outside} of the large demag. field zone) with the copper cell mounted (back: ``weakly-coupled'' drum run, front: this run). The $^3$He thermometer is also shown. }}
			\label{fig_6}
		\end{figure}
		
\subsection{Multiple runs}

3 different chips have been measured on the 2 platforms. They are mounted in copper cells with SMA connectors. On the nuclear demag. cryostat, the cell is bolted on a copper clamp that is located at the bottom of the laminar copper stage; the $^3$He thermometer is just underneath it. The copper clamp is connected to the stage and the thermometer on either side with 1 mm diameter annealed silver wires. Pictures of the demag. cryostat are shown in Fig. \ref{fig_6}.

Of the 3 samples measured, we have: a beam NEMS made in Grenoble which was used in Ref. \cite{ZhouPRAppl}, a drum-head NEMS made in Aalto and used in Ref. \cite{CiteCattiauxPRR} that we shall call ``weakly-coupled'', and another drum-head from Aalto (the one of this experiment), that we shall call ``strongly-coupled''.
The beam device displayed in the first flexure large amplitude fluctuations (nicknamed ``spikes'') at low temperatures, described in Ref. \cite{ZhouPRAppl}. They come from a stochastic driving force of unknown origin, that has been observed in different laboratories over the world. The first drum (``weakly-coupled'') showed below about 30$~$mK an apparent thermal decoupling of the fundamental mode that seemed to share some of the ``spikes'' characteristics \cite{CattiauxThese}. It was therefore impossible to measure in thermal equilibrium the mechanical modes of these structures down to (and below) 10$~$mK. For both devices, the microwave power required for the measurements is quite large, $n_{cav}$ in the range $10^4 - 10^6~$photons confined in the cavity (see Section \ref{microwave} below for the definition of $n_{cav}$).

However, the second drum could be measured down to very low powers, typically $300 - 600~$photons (using the so-called blue and red-detuned schemes, see below), thanks to its ``strongly-coupled'' nature. In these conditions, none of the above problems was seen (consistently with results from other labs, e.g. Ref. \cite{Schwab7mK}); the remaining issues are fluctuations in the device properties themselves, which are always seen in nanomechanical experiments (as far as they are looked for), as discussed in the core of the paper (and see below, section \ref{statistics}). The link between fluctuations, ``spikes'', and microwave power remains mysterious, and one can only speculate on that (see Ref. \cite{ZhouPRAppl} and Section \ref{statistics} below).

\section{Single-tone microwaves}
\label{microwave}

\subsection{Chip properties}

The device studied in this experiment is an aluminium drum-head (see Fig. \ref{fig_1} picture) of diameter 15$~\mu$m, 100$~$nm thickness suspended about 50$~$nm above a gate electrode coupled to a $\omega_c = 2 \pi \times 5.7~$GHz meander microwave cavity. It is made on sapphire. The total damping of the cavity resonance is $\kappa_{tot} = 2 \pi \times 500~$kHz, for an external coupling of $\kappa_{ext} = 2 \pi \times 240~$kHz. The measurement is performed in reflection, with a single coupling port; we are in the resolved sideband regime $\omega_m \gg \kappa_{tot}$.

The two lowest modes of the mechanical structure have been measured; the strongly coupled $[0,0]$ fundamental mode at $\omega_m = 2 \pi \times 15.1~$MHz (for a low temperature damping of $\Gamma_m = 2 \pi \times 420~$Hz), and the first non-axisymmetric mode $[0,1]$ at 25.9$~$MHz (for a damping of about 100$~$Hz) with a very small optomechanical coupling (non-zero because of the slight imperfect mode shape due certainly to the exact clamp structure). The next axisymmetric mode $[1,0]$ should be around 36$~$MHz but has not been characterized. From simulations, the device stores about 240$~$MPa stress.
Mechanical damping is dominated by (rather large) clamping losses below 100$~$mK, which is certainly linked to the large amount of stress present in the device \cite{ignacio}. 
In our experiment, the relaxation time of the mode is thus $\tau_m = 2/\Gamma_m \approx 1~$ms, independent of temperature below 100$~$mK.

\subsection{Basic calibrations}

\begin{figure}[t!]		
			 \includegraphics[width=19.5cm]{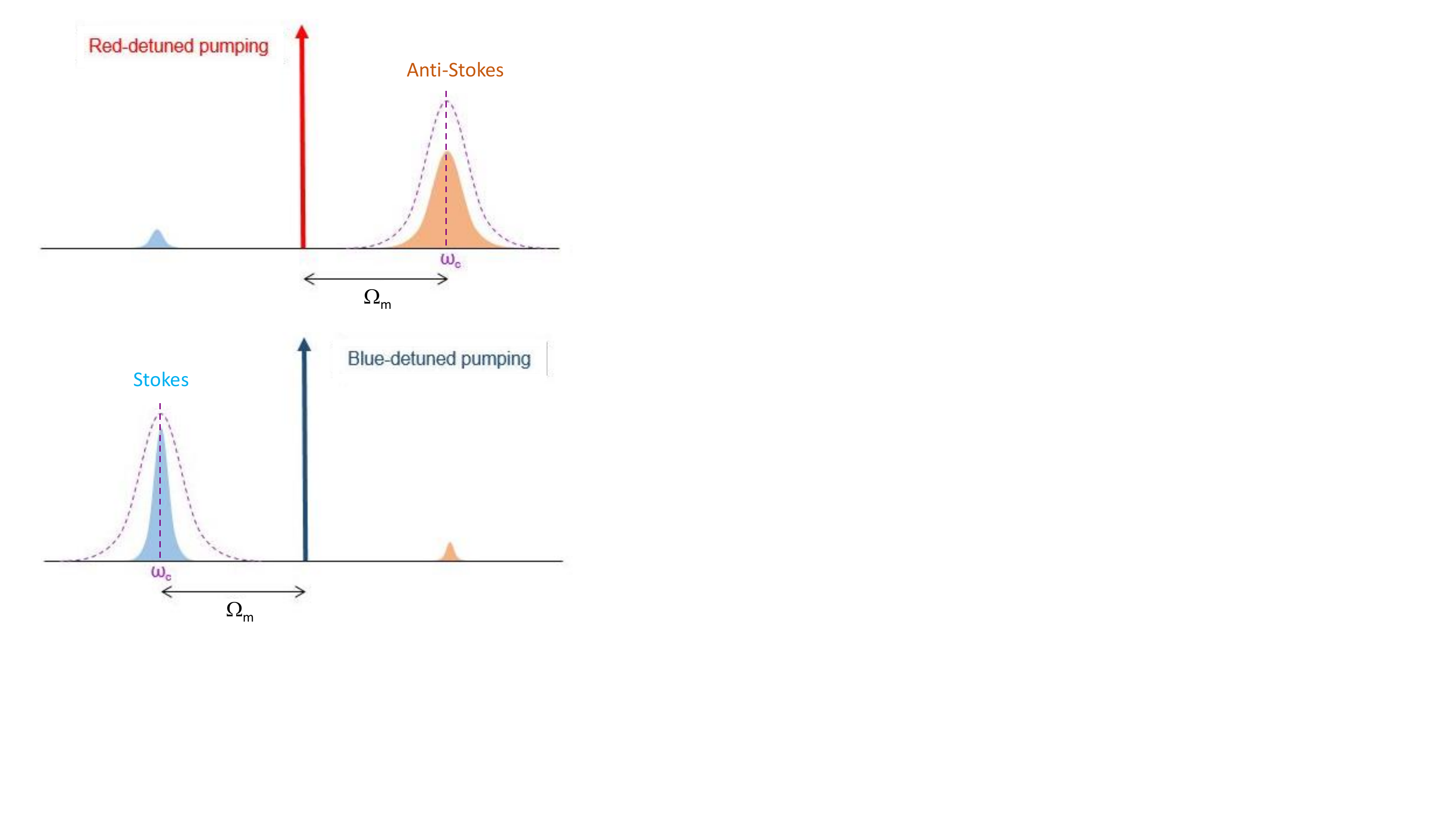}			
\vspace*{-3.cm}
			\caption{\small{  {\bf Optomechanics schemes.} Output spectra, with the cavity response depicted as dashed (not to scale). The pump tone is the vertical arrow (Dirac peak). Top: red-detuned pumping scheme. Bottom: blue-detuned pumping scheme. Only the enhanced sideband at $\omega_c$ is measured.}}
			\label{fig_6bis}
		\end{figure}
		
Only strict single-tone optomechanics has been used, such that the basic theory does strictly apply. This means that blue (microwave drive frequency $\omega_c+\omega_m$) and red (microwave drive frequency $\omega_c-\omega_m$) detuned pumping experiments have been performed in distinct runs. Only these two schemes have been used; see Fig. \ref{fig_6bis} for a schematic of the techniques.
The ``green'' pumping (pump tone applied at the cavity frequency $\omega_c$) could not be used; below typically 50$~$mK, the areas of the two measured sidebands were not consistent anymore with the cryostat temperature, a feature that is observed with the ``spike'' problem \cite{ZhouPRAppl}. We suspect that this is again due to the large amount of microwave power applied (which for this scheme is again on the order of $10^5$ photons). This feature, which seems to affect all microwave optomechanical devices, deserves to be studied but is outside of the scope of this work. \\

\begin{figure}[h!]		 
			 \includegraphics[width=11.5cm]{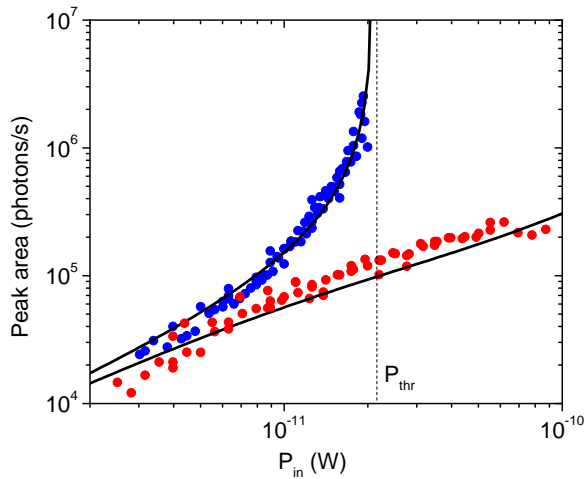}		
				\vspace*{-2.3cm}
			\caption{\small{  {\bf Area of sideband versus power.} Blue and red detuned measurements performed at 100$~$mK. Lines are fits (a slight out-of-equilibrium contribution is visible at large powers for red detuned pumping, which is discussed later in the text); the dashed vertical marks the threshold towards self-sustained oscillations (see text).  }}
			\label{fig_7}
		\end{figure}

\begin{figure}[t!]		 
			 \includegraphics[width=11.5cm]{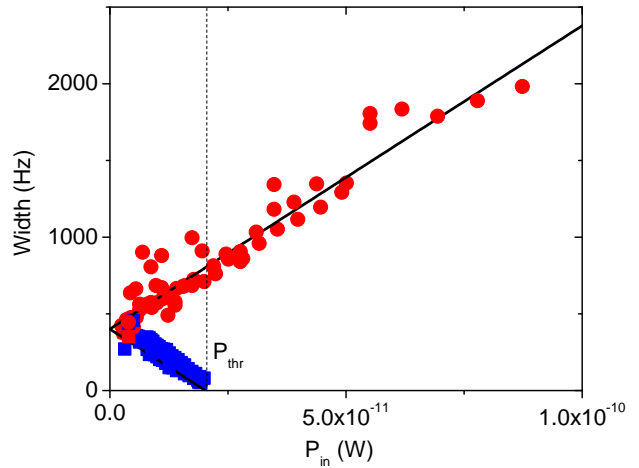}		
				\vspace*{-2.cm}
			\caption{\small{  {\bf Width of sideband versus power.} Blue and red detuned measurements performed at 100$~$mK. Lines are fits (pointing at $\Gamma_m$ for zero power injected); the dashed vertical marks the threshold towards self-sustained oscillations (see text). }}
			\label{fig_8}
		\end{figure}

We first characterized the optomechanical coupling $g_0$ by measuring at fixed temperature $T_{cryo} = 100~$mK the Brownian sideband peak as a function of applied microwave power. 
The peak areas and the peak widths are shown in Figs. \ref{fig_7} and \ref{fig_8} respectively. 

The lineshapes are Lorentzians (see fits Fig. \ref{fig_2} a insets). Basic formulas allow to link the measured peak area $A$ (in photons/s) and linewidth $\Delta W$ (in rad/s) to the effective NEMS mode population detected $n_{e f \! f}$ and applied microwave power $P_{in}$ \cite{AKMreview, ZhouPRAppl, CattiauxThese}:
\begin{eqnarray}
A & = & \Gamma_{opt} \,  n_{e f \! f} \,\,\,\,\,\,\,\,\,\,\,\,\,\,\,\,\,\,\,\, \mbox{for red}, \label{eq1} \\
A & = & \Gamma_{opt} \left(  n_{e f \! f} +1 \right)                  \,\,\,\, \mbox{for blue}, \label{eq2} \\
\Delta W     & = & \Gamma_m \pm \Gamma_{opt}, \\
\Gamma_{opt} & = & 4 \frac{g^2}{\kappa_{tot}}, \\
g^2          & = & g_0^2 \, n_{cav} , \\
n_{cav}      & = & \frac{P_{in} \, \kappa_{ext}}{ \hbar \omega_c \, \omega_m^2}, \\
n_{e f \! f} & = & \frac{n(T) \, \Gamma_m + N_{noise} \Gamma_{opt}}{\Gamma_m \pm \Gamma_{opt}}, \label{eqnoise}\\
n(T) & = & \frac{1}{Exp[(\hbar \omega_m)/(k_B T)]-1},
\end{eqnarray}
with the sign referring to red (+) or blue (-) detuning schemes, 
 using the resolved-sideband condition $\kappa_{tot} \ll \omega_m$. $n(T)$ is the Bose-Einstein thermal distribution, where $T=T_{mode}$ by definition. $n_{cav}$ is the number of drive photons confined in the cavity for the measurement. From fits (black lines) we extract the single phonon coupling $g_0 = 2 \pi \times 230~$Hz.
	
Let us discuss these simple expressions and what they actually mean. At high temperature when $n_{e f \! f} \gg 1$, Eqs. (\ref{eq1},\ref{eq2}) tell us that the areas obtained with blue and red pumping schemes are essentially equivalent; they are just proportional to the mode population $n_{e f \! f}$. However, when $n_{e f \! f}$ becomes very small, they start to be different: this effect known as sideband asymmetry \cite{PRXSchwab}, is in itself a proof that the system is very cold.
In Eq. (\ref{eqnoise}) which defines the effective mode population, we see that an extra term appears in addition to the Bose population: $N_{noise}$.
This term is linked to the photon noise $N_{cav}$ present in the microwave cavity (not to be confused with $n_{cav}$) through:
\begin{eqnarray}
N_{noise} & = & N_{cav} \,\,\,\,\,\,\,\,\,\,\,\,\,\,\,\,\,\,\,\, \mbox{for red},  \label{eqnoiseRed} \\
N_{noise} & = &   N_{cav} + 1                  \, \,\,\,\,\,\,\,\,\, \mbox{for blue}. \label{eqnoiseBlue}
\end{eqnarray}
For simplicity in our discussion, we assumed that the cavity photon noise and the one arising from the microwave port are the same.
In Eq. (\ref{eqnoiseRed}), we neglected a term $(\kappa_{tot}/4 \omega_m)^2$ which is small in the limit of sideband resolved setups; this is actually the lower bound for red detuned active cooling fixed by quantum mechanics \cite{AKMreview}.
A generic discussion can be found in Ref. \cite{CattiauxThese}.
The ``$+1$'' appearing in Eq. (\ref{eqnoiseBlue}) is a signature of zero point fluctuations in the cavity field \cite{AKMreview}.
If the noise $N_{cav}$ was limited only by the thermal noise of the cavity, and if we take the expected electronic temperature of 40$~$mK to evaluate it (see Section \ref{thermal}), we would obtain $10^{-3}$. This is clearly negligible, but unfortunately not the whole story, as shall be discussed in Section \ref{techheat}.

\begin{figure}[h!]		 
			 \includegraphics[width=15cm]{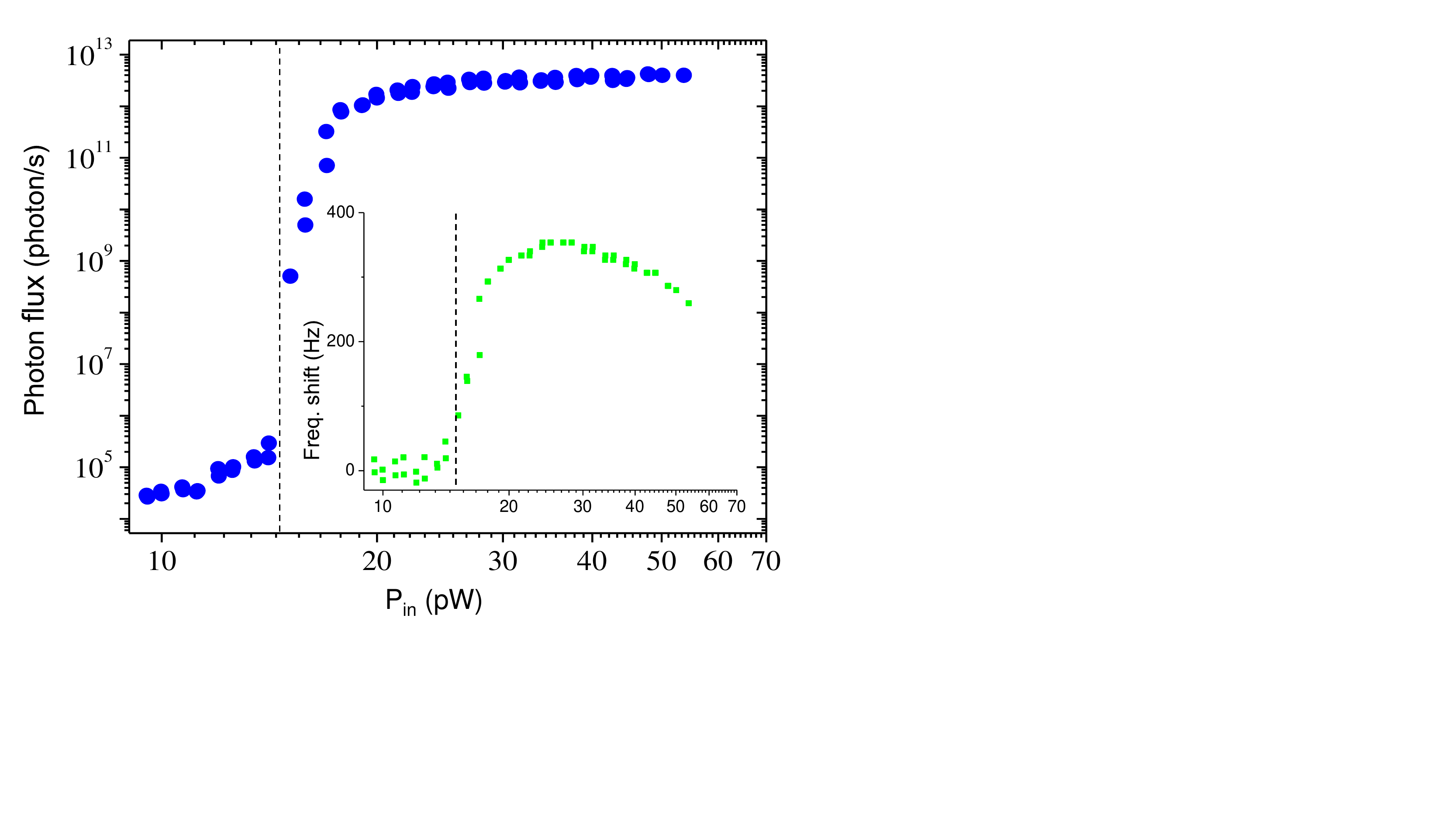}	
					\vspace*{-2cm}
			\caption{\small{  {\bf Self-sustained oscillations.} Amplitude of sideband signal as a function of power at 18$~$mK, in the blue detuned scheme. The dashed line marks the threshold were the linewidth becomes zero (see Figs. \ref{fig_7} and \ref{fig_8}). Inset: measured mechanical frequency shift versus power. }}
			\label{fig_9}
		\end{figure}

The red detuned scheme de-amplifies/cools the mode (population decreased, linewidth increased) while the blue scheme amplifies/heats (population increased, linewidth decreased). When the linewidth becomes zero (at $P_{thr}$, Figs. \ref{fig_7} and \ref{fig_8}), the mechanical mode enters self-sustained oscillations \cite{CiteCattiauxPRR}. This is shown in Fig. \ref{fig_9}.
The amplitude of motion becomes then very large, and the mechanical frequency shifts because of the Duffing nonlinear effect (stretching of the membrane, Duffing coefficient about 20$~$Hz/nm$^2$, see inset) \cite{CattiauxDuffing}. From simulations \cite{CiteCattiauxPRR}, we infer that the motion is in the {\it nanometer} range. 
			
\begin{figure}[h!]		 
			 \includegraphics[width=16cm]{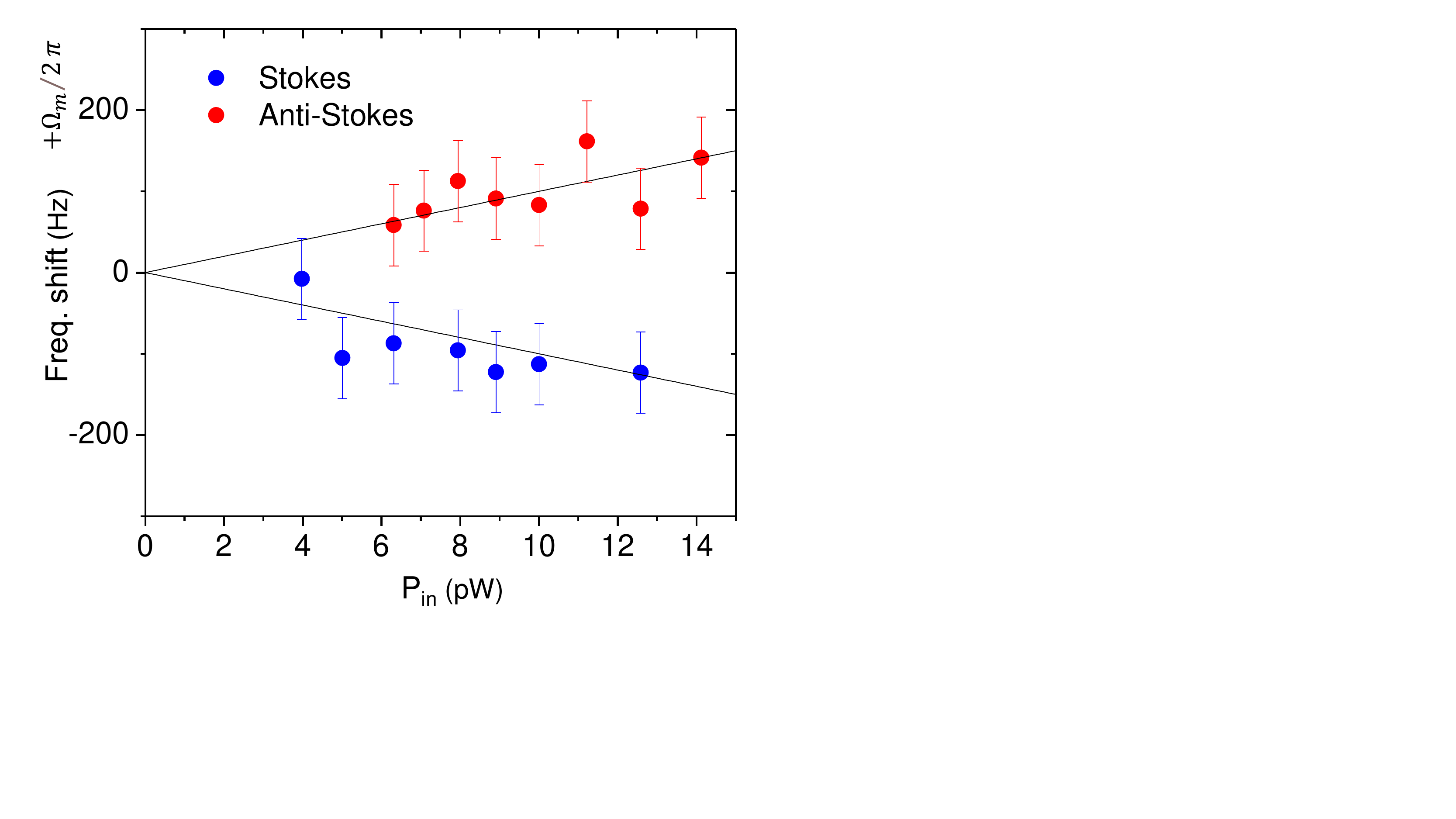}	
					\vspace*{-2.5cm}
			\caption{\small{  {\bf Optical spring.} Mechanical frequency shift as a function of microwave power at 15$~$mK for both red and blue detuned schemes. Lines are fits (see text). }}
			\label{fig_10}
		\end{figure}

When the pump tone is not perfectly tuned, the position of the measured sideband moves linearly with applied power; an effect known as optical spring \cite{AKMreview}. This is shown in Fig. \ref{fig_10} for our experimental settings, and explains the slight difference in the actual peak positions in Fig. \ref{fig_2} a (but has a marginal impact on other measured parameters). 
The pump position $\omega_c \pm \omega_0$ has been kept constant over the whole experiment (with $\omega_0$ the high temperature value of $\omega_m$). As such, the optical spring measured was temperature-independent; the $T$-dependent mechanical frequency shift (Fig. \ref{fig_1}) being far too small compared to $\kappa_{tot}$ to have any effect on this setting. 
But precisely because the mechanical frequency shifts with temperature, any microwave-induced physical heating of the device would add-up and create an additional ({\it and} temperature-dependent) contribution to the effect presented in Fig. \ref{fig_10} (see e.g. Ref. \cite{ZhouPRAppl}). This is not observed in our case, and we therefore conclude that {\it no} relevant microwave-induced physical heating of the device is present, at any temperature, in the range that we explored. 

The point of the experiment is to use optomechanics as a probe, as little invasive as possible: ideally for extremely small injected power $n_{cav} \rightarrow 0$, the amplification/de-amplification is negligible. Ideally as well, the stochastic back-action arising from the (very small, but) finite cavity/microwave port population $ \Gamma_{opt} \, N_{noise}$ should also be essentially zero. In this limit, one would just measure a sideband signal proportional to injected power, subject only to sideband asymmetry, i.e. obtaining an area from Eqs. (\ref{eq1}-\ref{eq2}) proportional to $n(T)$ in the red detuned scheme, and $n(T)+1$ for blue. This leads us to define the ``effective temperatures'' $T_{blue}, T_{red}$ from the area $A$ quoted in the core of the paper:  
\begin{eqnarray}
T_{blue}&=&(n+1)\frac{\hbar \omega_m}{k_B}                    \,\,\,      \mbox{blue-detuned }, \\
T_{red}&=&(n)  \frac{\hbar \omega_m}{k_B} \,\,\,\,\,\,\,\,\,\,\,\,\,\,   \mbox{red-detuned}.
\end{eqnarray}
These are practical parameters for the experimentalist: above typically 10$~$mK, they {\it both equalise} to the mode temperature $T_{mode}$. 
Having them follow $T_{cryo}$ is thus a proof of thermalisation. 

\begin{figure}[t!]
	\centering		 
			 \includegraphics[width=14.5cm]{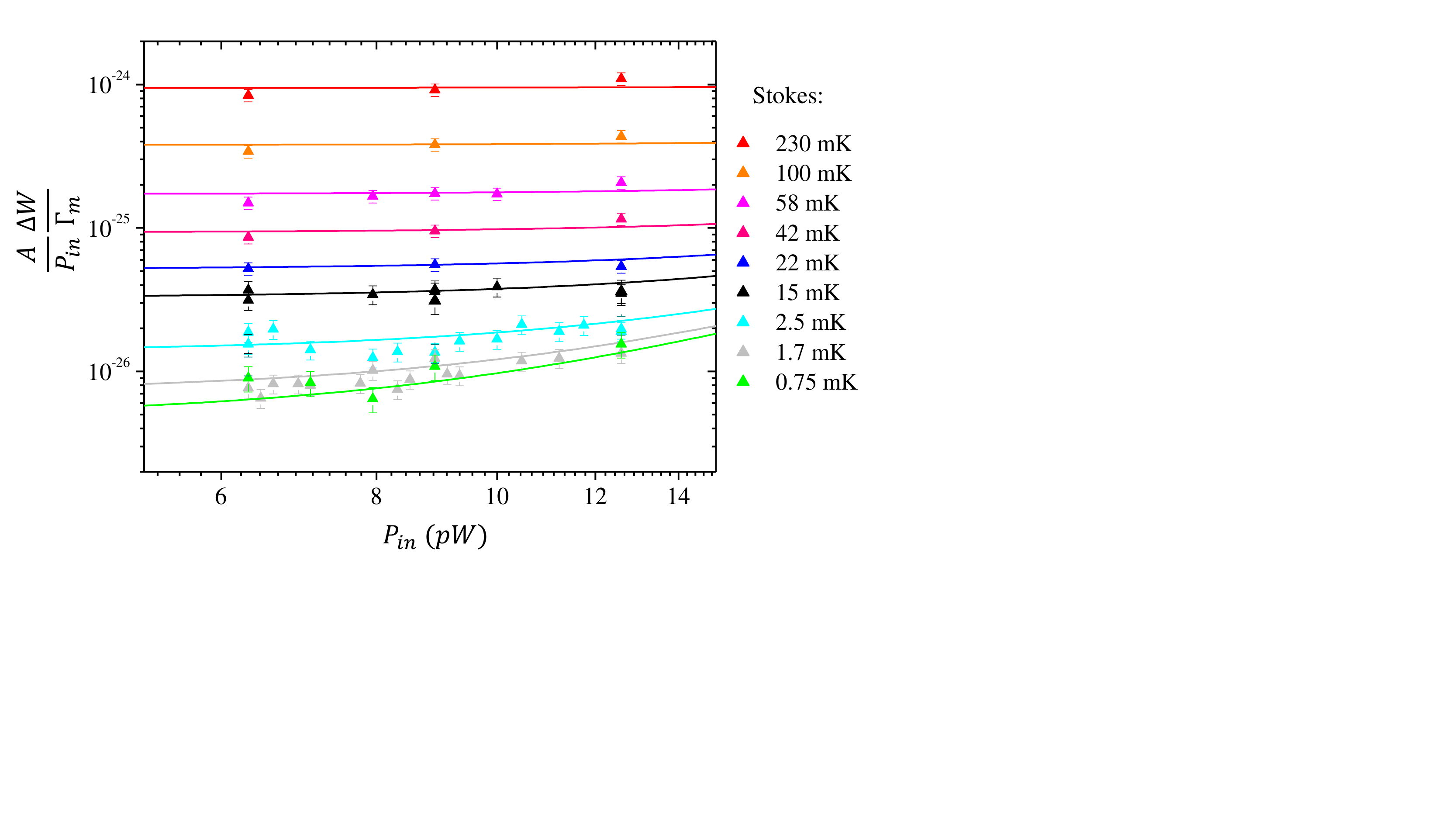}		\vspace*{-2.5cm}	 	
				
				\includegraphics[width=14.5cm]{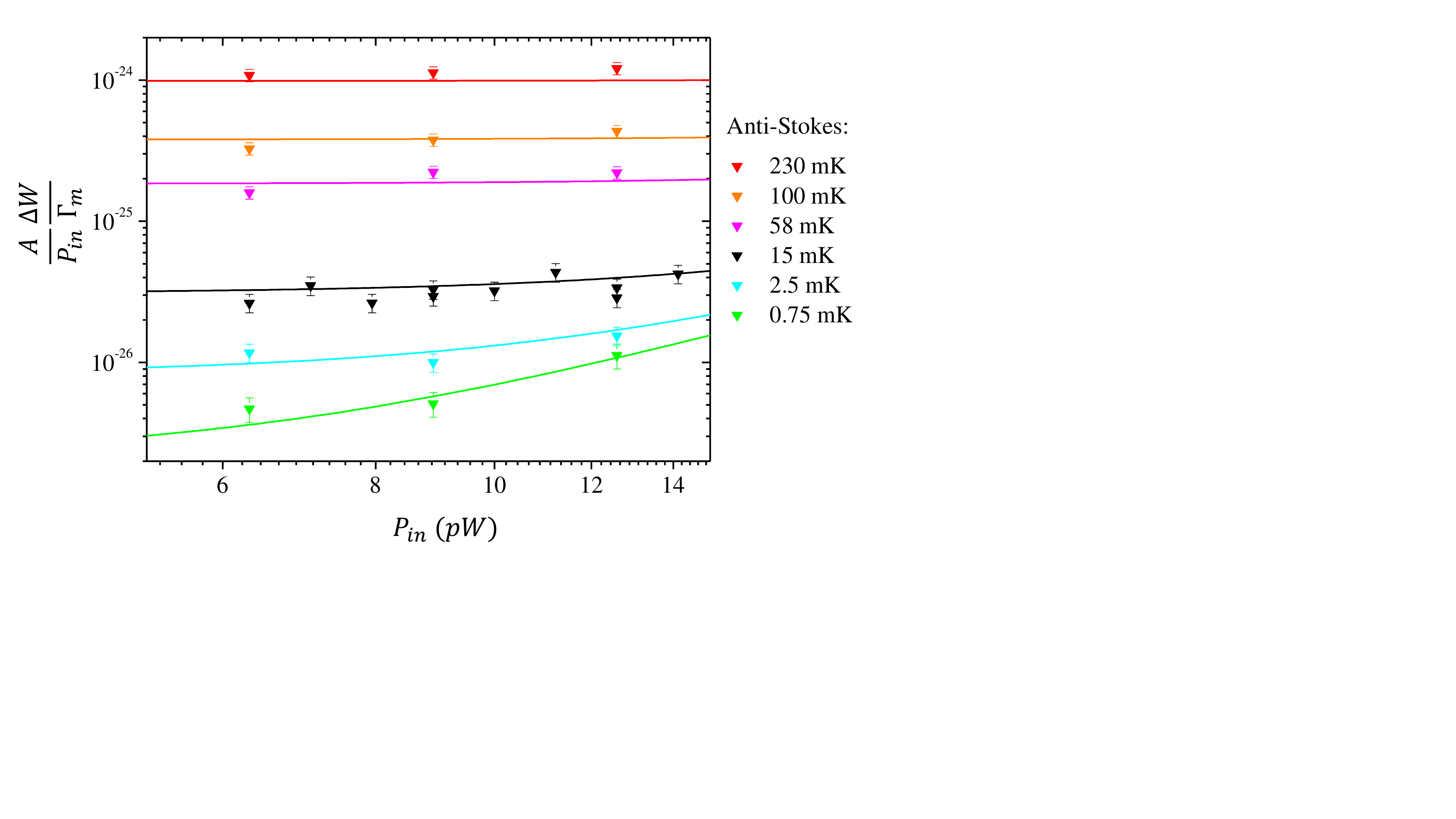}  \vspace*{-2.5cm}
			\caption{ \small{  {\bf Normalised areas.} Areas normalized to injected power and linewidth as a function of $P_{in}$. Top: blue-detuned pumping. Bottom: red-detuned pumping. Lines are fits explained in the text. }}
			\label{fig_11}
		\end{figure}

We therefore keep the drive power in the range 300 to 600 photons for $n_{cav}$, the lowest we can afford for practical reasons. Even if small there is still a measurable amplification/de-amplification in this range; but this is easily corrected for since the ratio $\Gamma_{opt}/\Gamma_{m}$ is known (see however the discussion in Section \ref{statistics} about damping fluctuations). 
This is what is applied to create the main graph of Fig. 2 a. However, the sideband peaks displayed in insets have not been corrected; this is why an asymmetry is visible at high temperature between the Stokes and anti-Stokes peaks.
As such, the ratio of their areas is bound at high temperature by   $(\Gamma_m-\Gamma_{opt})/ (\Gamma_m+\Gamma_{opt}) \approx 0.5$, as can be seen in the top inset, Fig. \ref{fig_2} a (sideband asymmetry thermometry). 
The line in the same inset graph is theory; in principle an independent proof of thermalisation down to the lowest temperatures. 

\subsection{Technical heating}
\label{techheat}

Unfortunately, the second assumption discussed in the previous paragraph 
(ideally cold microwave mode, $N_{cav}=0$)
 is also not fully verified: as we increase the injected microwave power $P_{in}$, there is {\it technical heating} of the 
microwave
mode (to be distinguished from physical heating of the whole structure; see Fig. \ref{fig_7} the red detuned data). This effect is known in the community, and not always easy to circumvent.
 It is presumably 
due to out-of-equilibrium photons that populate the cavity and act (through back-action) onto the mechanical degree of freedom. 
These out-of-equilibrium photons are believed to originate in the {\it phase noise} of the microwave source; an ultimate limitation of any setup. 
The foot of the Dirac peak of the pump ``leaks'' into the cavity, so to speak.
But the actual mechanism behind this is not fully understood, and the magnitude of this effect does depend on sample and cool down: it has to be carefully characterized in each run.

To demonstrate this for our experiment, we plot in Fig. \ref{fig_11} the area $A$ normalised to power $P_{in}$  and linewidth $\Delta W$ as a function of $P_{in}$, for different temperatures $T_{cryo}$. 
And indeed, we see at the largest power a deviation, characteristic of this effect (without which it should remain flat, and $\propto T_{blue}, T_{red}$). It can be fit by an out-of-equilibrium contribution $\propto (P_{in})^a$ (a power law dependence with exponent $a$) \cite{ZhouPRAppl}, see lines in Fig. \ref{fig_11}. This term is independent of the scheme, and of temperature. It is of the order of 1 out-of-equilibrium photon at the lowest drive used (300 photons), and can be easily corrected for; 
it is negligible in all conditions presented in the core of the paper, except for two measurements performed around 700$~\mu$K (the time-trace taken with the blue scheme, Fig. \ref{fig_2} b, 600 photons drive, and the anti-Stokes peak Fig. \ref{fig_2} a, 300 photons), and the time-trace taken at 1.4$~$mK (blue scheme, Fig. \ref{fig_2} b, 600 photons drive). In these extreme cases, the correction nonetheless remains smaller than a factor two.
In order to certify that there is no flaw in the analysis, a time trace with 300 photons has also been taken and processed at 14$~$mK; and indeed results are identical for 300 and 600 drive photons (see Section \ref{statistics} below).

As a concluding remark on the mode populations, we can compute the $n(T)$ of the nearest modes at the lowest temperature of 500$~\mu$K assuming perfect thermalization, which as discussed in the core of the paper is essentially guaranteed by our results obtained on the lowest mode (see also the discussion on thermal properties in Section \ref{thermal} below). 
At this temperature the fundamental mode $[0,0]$ stores 0.3 quanta on average. The nearest non-axisymmetric mode $[0,1]$ stores about 0.1 quanta while the next axisymmetric mode $[1,0]$ has 0.03 quanta. All higher modes verify $n(T) \leq 1~\%$. 

\section{Statistical analysis}
\label{statistics}

Beyond thermalisation, a striking result of the experiment concerns the very slow, and rather large fluctuations that are seen in mechanical properties as a function of time.
Obviously our measurement is not single-shot (detection background about 100 photons at 5$~$GHz), and it is not a QND measurement of the population ($\hat{n}$ operator); what we measure is the spectrum of the $\hat{x}$ operator, the position.
We need between 5 and 20 minutes at our highest drive level in blue-detuned pumping (600 photons) in order to have an acquisition data averaged enough to be fitable (depending on temperature). This means that what we measure is a ``smoothed'' estimator of the time-dependent mechanical properties, but not the real-time ones. But it is an invaluable, and unique, information about whatever processes are at stake in the thermodynamical equilibrium of the device.

The setup and measurement scheme has been kept basic, so that it can be easily modelled 
by ``conventional single-tone optomechanics''. This makes the interpretation of measured data rather simple, and the only experimental parameter linked to the detection that we will have to consider for potential biases in the analysis is the (even if small) power $P_{in}$ (or number of photons $n_{cav}$). This has been already discussed in the previous Section. And of course, because of the nature of the data presented we have to consider carefully the potential flaws that the statistical analysis may imply. These aspects are the subject of this Appendix.

		\begin{figure}[h!]		 
			 \includegraphics[width=15cm]{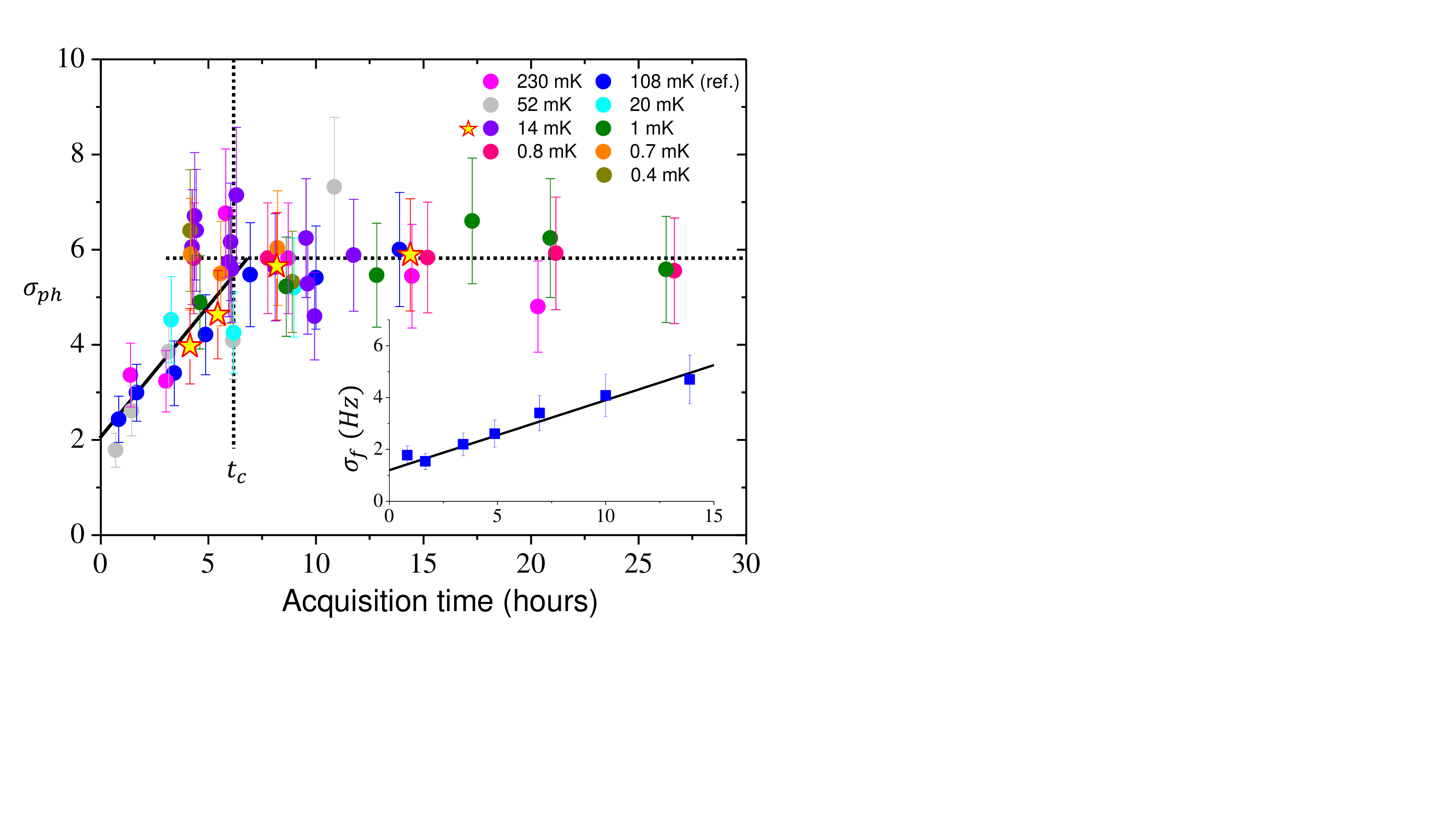}	
					\vspace*{-2cm}
			\caption{\small{  {\bf Standard deviation versus acquisition time.} Phonon standard deviation measured at different $T_{cryo}$ normalised to 108$~$mK data, with an averaging window of 20 minutes (from 600 photon drive in blue-detuned scheme). Inset: frequency standard deviation at 108$~$mK. The stars are measured with 300 photons (see text). }}
			\label{fig_12}
		\end{figure}	

\subsection{Potential statistical flaws}

The first statistical parameter we can consider is the full acquisition time. In Fig. \ref{fig_12} (main graph) we show the phonon standard deviation $\sigma_{ph}$ measured at different $T_{cryo}$ scaled on the 108$~$mK values, as a function of the length of the acquisition time. In inset, we give the frequency standard deviation $\sigma_f$ measured at 108$~$mK (damping noise is discussed in the next Section). For these data, all fits have been performed on time-traces where single points are obtained by 20 minutes averages. As explained in the core of the paper, this averaging window is sliding through the acquired set of data (which has a time step of 1 second). Apart from the highest temperatures, no peak can be seen on single files.
$\sigma_{ph}$ and $\sigma_f$ have rather different behaviours, which are characteristic of the nature of their fluctuation spectra. For full acquisition times smaller than $t_c=5$ hours, $\sigma_{ph}$ grows essentially linearly, similarly to the whole behaviour of $\sigma_f$. This comes from the $1/f^2$ nature of the spectra. However, above 5 hours it becomes flat: this is a signature of the low-frequency cutoff $1/t_c$ seen in the fluctuation spectrum. The standard deviation of the phonon noise is bound, and measuring with an acquisition time of 10 hours guarantees to have a proper estimate of $\sigma_{ph}$. On the other hand, $\sigma_f$ grows continuously: the spectrum is not bound at low frequency, this noise is actually {\it non-stationary}. We therefore have to quote for what acquisition length we compute $\sigma_f$; this is what people refer to in the literature with the Allan deviation \cite{NatNanotechHentz}. Here again, we shall compute the standard deviation with a time slot of 10 hours.

		\begin{figure}[h!]		 
			 \includegraphics[width=15cm]{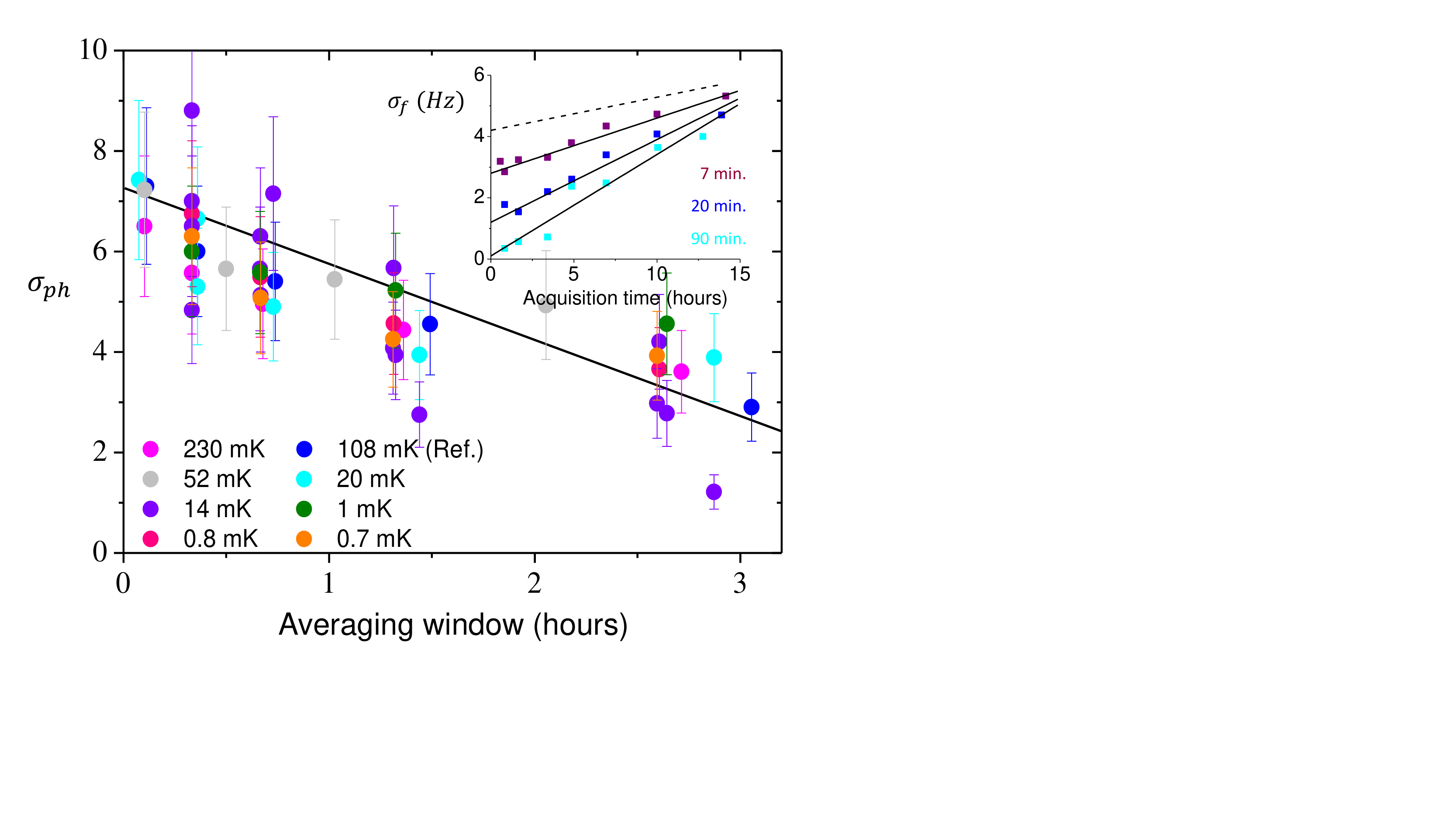}			
				\vspace*{-1.7cm}
			\caption{\small{  {\bf Standard deviation versus averaging window.} Phonon standard deviation at different $T_{cryo}$ for acquisition times longer than 10 hours scaled on 108$~$mK data (line is linear guide). Inset: frequency standard deviation versus acquisition time for 3 different averaging window, at 108$~$mK. Lines are fits, and the dashed one is the extrapolation for zero averaging (see text). All data taken at 600 photons in blue-detuned scheme. }}
			\label{fig_13}
		\end{figure}	

More importantly, we have to consider the effect of the averaging window on the quoted standard deviations. In Fig. \ref{fig_13} we show the effect of the size of the window on the $\sigma_{ph}$ measured in the plateau region of Fig. \ref{fig_12} (beyond 10 hours acquisition, scaled on the value at 108$~$mK). As expected, we see a rather smooth decrease of $\sigma_{ph}$ when we increase the averaging time. At first order, this looks like a linear dependence; at most, when averaging 20 minutes we lose about $20~\%$ of the total phonon fluctuation amplitude. This has been corrected for in Fig. \ref{fig_3} of the core of the paper. 
In inset on Fig. \ref{fig_13}, we demonstrate the impact of the averaging procedure on $\sigma_f$; we plot it as a function of acquisition time for 3 different averaging windows. As expected, the more we average the smaller the $\sigma_f$ value. In all cases, the dependence seems linear with respect to acquisition length. Increasing the averaging time seems to impact much more the short acquisitions (the 0-time intercept goes down very quickly by increasing the averaging), while it is less pronounced at large times. For 10 hours acquisitions, a 20 minutes averaging window decreases the impact of frequency fluctuations on the data by about $20~\%$ (see dashed line in Fig. \ref{fig_13}). This has been corrected for in the $\sigma_f(T)$ plots.

One important aspect of the measurement which also has to be discussed within the statistical analysis is the finite power at which (in blue-detuned pumping) the measurement is performed. For technical reasons, this amplifies the fluctuations: even when the average population is below one, the acquisition is performed around about 3 phonons. It is thus expected than, within our resolution all distributions shown in Fig. \ref{fig_2} b look Gaussian (in accord with the central limit theorem). Knowing the gain $\Gamma_m/(\Gamma_m-\Gamma_{opt})$, one can easily recalculate the original average value. However, it is perfectly reasonable to wonder if the procedure could modify the amplitude of fluctuations (like e.g. {\it squeeze} them). We therefore performed a time-trace measurement at half the drive (300 photons), at a temperature high enough such that the averaging settings we use would still be enough to process the data over decent timescales (we took 14$~$mK; see stars in Fig. \ref{fig_12}). This leads to the orange dot in the graphs (Fig. \ref{fig_3} of the paper, Fig. \ref{fig_14} below and stars in Fig. \ref{fig_12}). As one can clearly see, the obtained standard deviations are equivalent to the ones extracted with 600 photons. 
The overall shape of the spectra reported are also independent of microwave power and statistical analysis: the $1/f^2$ nature is absolutely robust (for both phonons and frequency/damping noises), as well as the $t_c$ cutoff time seen for phonon fluctuations.
We thus conclude that there is no flaw in the magnitude of the fluctuations reported. 
Also: 
 if there was an influence from other sources of noise (especially electric noise for what concerns the optical field, damping noise impacting the area measurement or fluctuations in cryostat temperature), we would expect an effective saturation at the lowest temperatures. This is not observed, and $\sigma_{ph}$ seems to follow $n$ monotonically in Fig. \ref{fig_3}.

In order to conclude our discussion of the statistical analysis, we shall comment on the {\it relevant timescales} involved in the experiment.
The physical filtering is about 1 Hz from the lock-in, and the averaging extrapolation is performed from a typical time of a minute.
Therefore, our analysis is a proper treatment of the data that reproduces genuinely {\it only the slow component} of the population fluctuations. 
To be concrete, the inset of Fig. \ref{fig_1} suggests that the mechanical damping writes $\Gamma_m = \Gamma_{ph}+ \Gamma_{TLS}$, with $\Gamma_{ph} \gg \Gamma_{TLS}$ (and thus $\Gamma_m \approx \Gamma_{ph}$).
The phonon contribution is by far too fast compared to our measurement; its contribution to the statistics is suppressed by a factor of the order of 10$~$s$/$1$~$ms. 
On the other hand, the TLS contribution $\Gamma_{TLS}$ (presumably of order $1/t_c$) is slow enough to be perfectly reproduced by our analysis: this is the genuine mechanism that has to underlie our results.
Also by construction, our data actually contain {\it all possible sources of slow noise} integrated together (in particular the impact of $1/f$-like electronic noise, the $1/f^2$ damping noise discussed below, and even the fridge long-term temperature instability). 
It is also true for the frequency fluctuations. One could therefore guess that our $\sigma_{ph}$, $\sigma_f$ could only possibly over-estimate genuine TLS- fluctuations, if any imperfection is to be claimed. 
For this reason, it is rather surprising to measure a sub-Poissonian statistics; it would have been naively more natural to expect a result above the $\sigma_{ph}=\sqrt{n}$ dependence. We thus conclude that the prefactor $0.5$ in the $\sigma_{ph}=0.5 \sqrt{n}$ law 
provides important insight into the TLS bath and the way in which it interacts with the mechanics.

\subsection{Frequency and {\it damping} fluctuations}

\begin{figure}[t!]		 
			 \includegraphics[width=14cm]{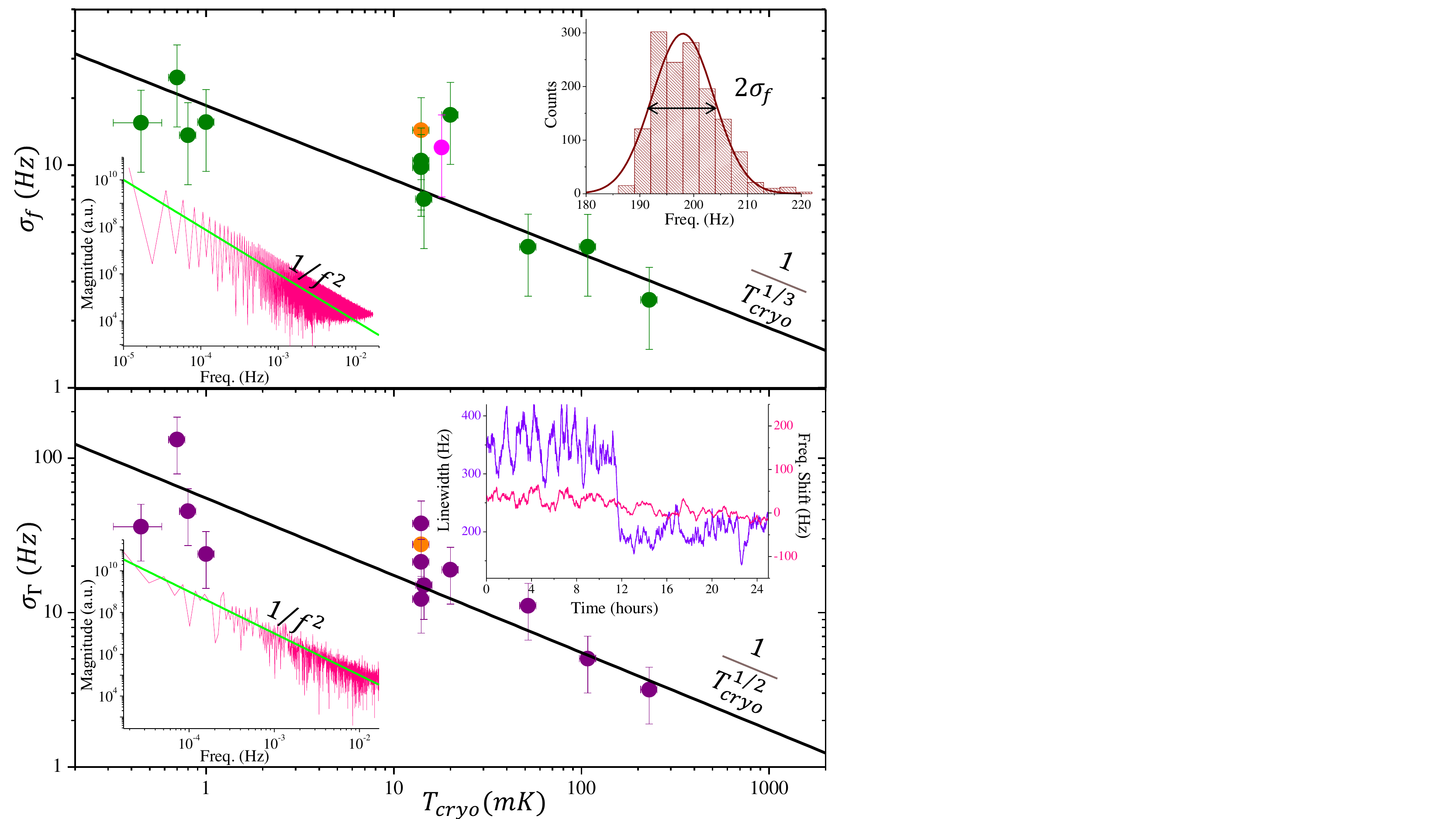}		
				\vspace{0.5cm}
			\caption{\small{  {\bf Damping noise.} Comparing frequency (top) and damping (bottom) fluctuations as a function of temperature (standard deviation $\sigma_\Gamma$ computed for 10$~$h acquisition, as for $\sigma_f$). The magenta point is inferred from the self-oscillating state measurements (see Fig. \ref{fig_9} above), in agreement with the Brownian motion result (also on Fig. \ref{fig_3} of the paper at 18$~$mK).
			Insets: Gaussian like statistics and $1/f^2$ spectra. In the last inset, we report on a damping sudden jump, while {\it nothing} is seen on the frequency. The black lines in the main graphs are power laws $1/(T_{cryo})^a$ guides to the eye. The orange dots, as in other graphs, are obtained at half the drive (see text). }}
			\label{fig_14}
		\end{figure}

Besides frequency and area fluctuations, we also report on {\it damping fluctuations}. These are usually not looked at, but are always present in nanomechanical systems \cite{MailletACSNano}. Actually, all the essential characteristics of the frequency noise ($1/f^2$ spectrum, Gaussian-like distribution, rare telegraph-like jumps, and $1/T^a$ temperature dependence below 200$~$mK) appear to be shared by damping noise. However, these two noises seem to be not (at least perfectly) correlated. This is demonstrated in Fig. \ref{fig_14}. 

Even if not strictly equivalent, the frequency noise looks quite similar to what is found in superconductors: $1/f^a$-like, growing with decreasing temperature \cite{burnett}.
Also, the damping noise seems to dominate over other sources of errors 
(namely the amplifier noise, and then the fit error)
when evaluating mechanical parameters, and is thus responsible for the scatter in the data. 
We can only speculate on the origin of all material-dependent fluctuations, namely ``spikes'' \cite{ZhouPRAppl}, frequency noise, and damping noise. These may be related to the same underlying mechanisms, but it will require further work to be proven. Two-Level-Systems (TLS) are often invoked when dealing with these issues, since they explain the frequency shifts $\omega_m(t)-\omega_0$ as well as the high temperature dependence of the damping $\Gamma_m$.
The exact fit function (dashed line in Fig. \ref{fig_1}) for the frequency shift 
 reads $\omega_m (T)-\omega_0 \propto (Re[\Psi(1/2+(\hbar \omega_0)/(2 \pi i k_B T))]-Ln[(\hbar \omega_0)/(k_B T)])$ with $\Psi$ the Digamma function \cite{ReviewPhilips}. This leads to the observed logarithmic-like dependence (full line in Fig. \ref{fig_1}).
The damping happens to be linear with temperature at high temperature, a feature also observed for other NEMS devices in the same temperature range \cite{OlivePRBTLS}. 
An intriguing effect has been noticed in the mechanical frequency shift at the lowest temperatures (below 10$~$mK): even though the equilibrium temperature follows the Log (or equivalently Digamma) law, we discovered temperature hysteresis when ramping up the temperature after cooling. 
This could be the signature of some sort of ``glassy'' behaviour within the TLS bath (subject to TLS-TLS interactions? See e.g. Ref. \cite{AndrewParpiaTLS,burnett}), but we are again limited to speculations and one would need further investigations to understand this.
More discussions can be found in Ref. \cite{CattiauxThese}.



\section{Thermal modeling}
\label{thermal}


The experimental work presented in the paper is about the thermalisation of a mesoscopic mechanical object down to 500$~\mu$K. 
We present in this Appendix the basic thermal modeling of the device, that consistently supports our experimental result.
First, we will review the fundamental models to calculate the specific heat (Debye model) and the thermal transport (Kinetic equation and Casimir model).
Basic material-dependent parameters will be reminded.\\

Nomenclature:
\begin{itemize}
	\item $k$ thermal conductivity W.m$^{-1}$.K$^{-1}$
	\item $K$ thermal conductance W.K$^{-1}$
	\item $c_p$ specific heat in J.m$^{-3}$.K$^{-1}$ 
	\item $C$ heat capacity J.K$^{-1}$
	\item $\Lambda$ phonon mean-free-path 
\end{itemize}

The modeling is kept basic, since only orders of magnitudes are sought. 
As such, we should point out already that the specific heat of materials is not affected by small dimensions at the 100-nanometer scale (we can thus keep using bulk values) \cite{Cleland}.
On the other hand, for the thermal conductance we {\it need } to take into account the small dimensions of the heat conductor (and thus renormalize the bulk values and adapt them to our 2D-membrane sample).

\subsection{Thermal properties of superconducting aluminium membrane}

\begin{figure}
	\begin{center}
		\includegraphics[width=11cm]{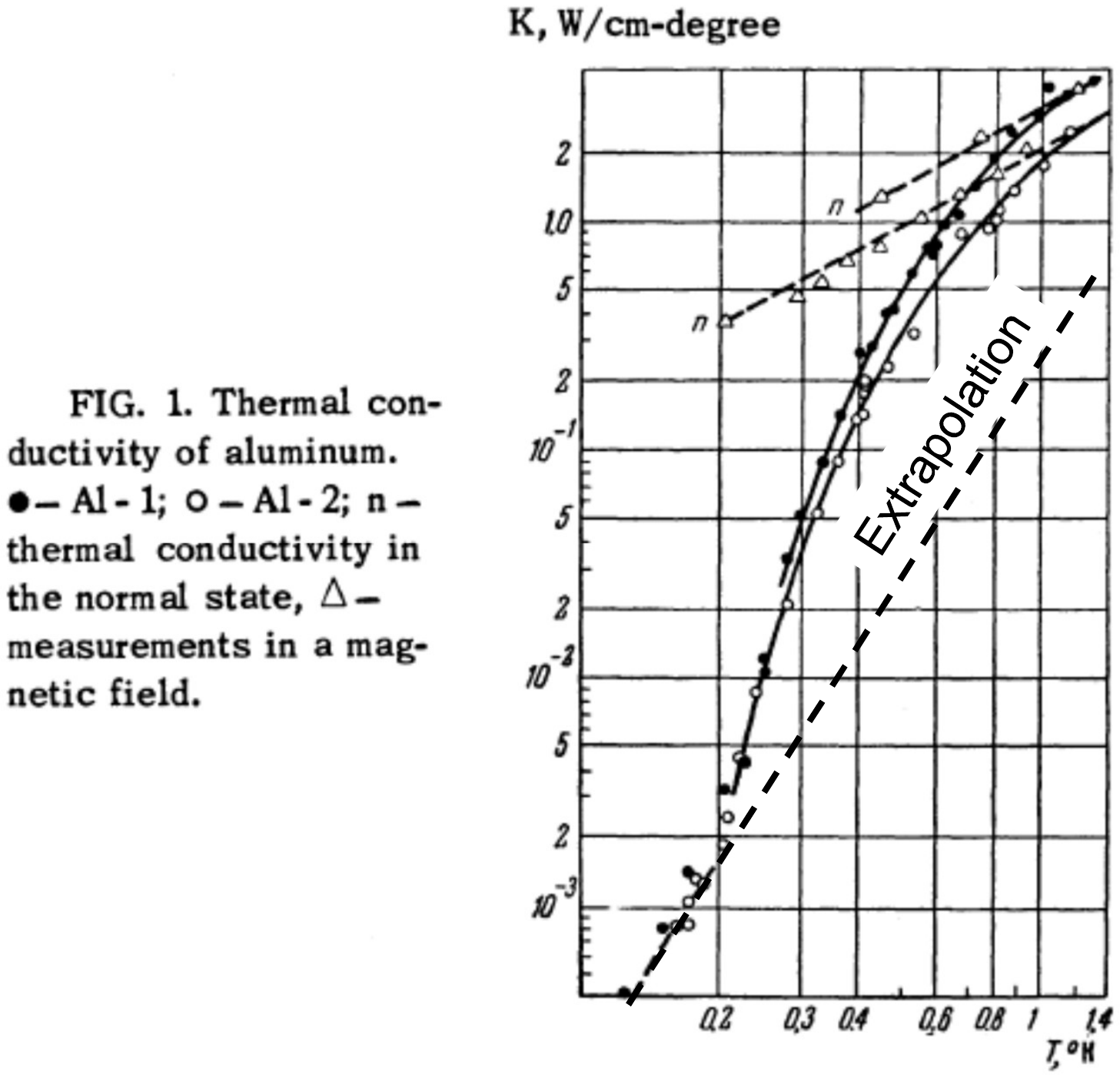}
		\vspace*{-2cm}
	\end{center}
	\caption{{\bf Thermal conductivity.} Measurement done by Zavaritskii \cite{zav} in 1958 of thermal conductivity in aluminium, on bulk materials (few centimeter large). The extrapolation of that curve and the corresponding fit are given in Fig.~\ref{fig3} (and Tab. \ref{table1}). Image adapted from Ref. \cite{zav}. }
	\label{fig1}
\end{figure}

The heat carriers that will matter for calculating the characteristic thermalization time (and other parameters) are the phonons. Indeed, thermal transport in {\it Al} at low enough temperature is dominated by phonons and not any more by electrons. 
The experiment is performed below 200$~$mK: the temperature is then far lower than the superconducting critical temperature $T_c$ of aluminium ($T_c \approx 1.2$~K); we can thus consider that all the conducting electrons are BCS-condensed, and consequently the thermal transport will happen through the lattice only (phonons), because the condensed electrons cannot carry entropy (then do not contribute to thermal transport) \cite{BCS}.
 This can be seen in the thermal transport measurements done by Zavaritskii \cite{zav} presented in Fig.~\ref{fig1} where  the thermal conductivity dramatically changes between $T_c$ and $T_c/5$: below 0.2~K one sees that only the phonon contribution remains as it varies like $T^3$.

As well, the contribution of electrons to the specific heat $c_p$ is decreasing exponentially below $T_c$; this is a standard result that can also be calculated in BCS-theory \cite{BCS}. We can thus also neglect electrons in the specific heat, and consider only lattice vibrations for the heat capacity of the object.
We shall come back at the end of this Appendix to electronic properties. 

\begin{table}
	\centering
	\begin{tabular}{|c|c|c|c|c|c|}
		\hline
		Material &  $ v_{s}$  &  $\theta_{D}$ & $\rho$    & $c_p$  & $k_{bulk}$\\
		\hline
		{\it Al} &  6700 &  468 &  2\,700  &    0.41 $\times T^{3}$&  23.4 $\times T^{3}$ \\
			\hline
			\end{tabular}
	\caption{Aluminium ({\it Al}) data for thermal properties: the speed of sound $ v_{s}$ in m/s, the mass density $\rho$ in kg/m$^3$, the Debye temperature $\theta_{D}$ (in K). 
	Specific heat in J.m$^{-3}$.K$^{-1}$ and (bulk) thermal conductivity in W.m$^{-1}$.K$^{-1}$. }
	\label{table1}
\end{table}

\subsection{Specific heat of the Al membrane}

We will use the Debye model to calculate the phonon specific heat of aluminium associated to the crystal lattice, in other terms the phonon specific heat. For $T\ll \theta_{D}$ (the Debye temperature), $c_p$ is given by the following equation using the averaged speed of sound $v_{s}$: 
\begin{equation}
c_{p}=\frac{2 \pi^2}{5}\frac{k_{B}^4}{\hbar^3 v_{s}^3} T^3.
\label{Debyevs}
\end{equation}
The specific heat is in J.m$^{-3}$.K$^{-1}$. To convert it to J.kg$^{-1}$.K$^{-1}$, if necessary, one has to divide Eq.~(\ref{Debyevs}) by the density $\rho$ in kg.m$^{-3}$ (see e.g. Ref. \cite{Cleland}).
Parameters are given in Tab. \ref{table1} with the thermal conductance fit on Ref. \cite{zav} (see Fig.~\ref{fig3}). We find a specific heat for the aluminium crystal lattice of $c_p = 0.41 \times T^{3}$ J.m$^{-3}$.K$^{-1}$. 

\subsection{Thermal conductivity, Kinetic equation and phonon mean-free-path in the structure}

\begin{figure}
	\begin{center}
		\includegraphics[width=10cm]{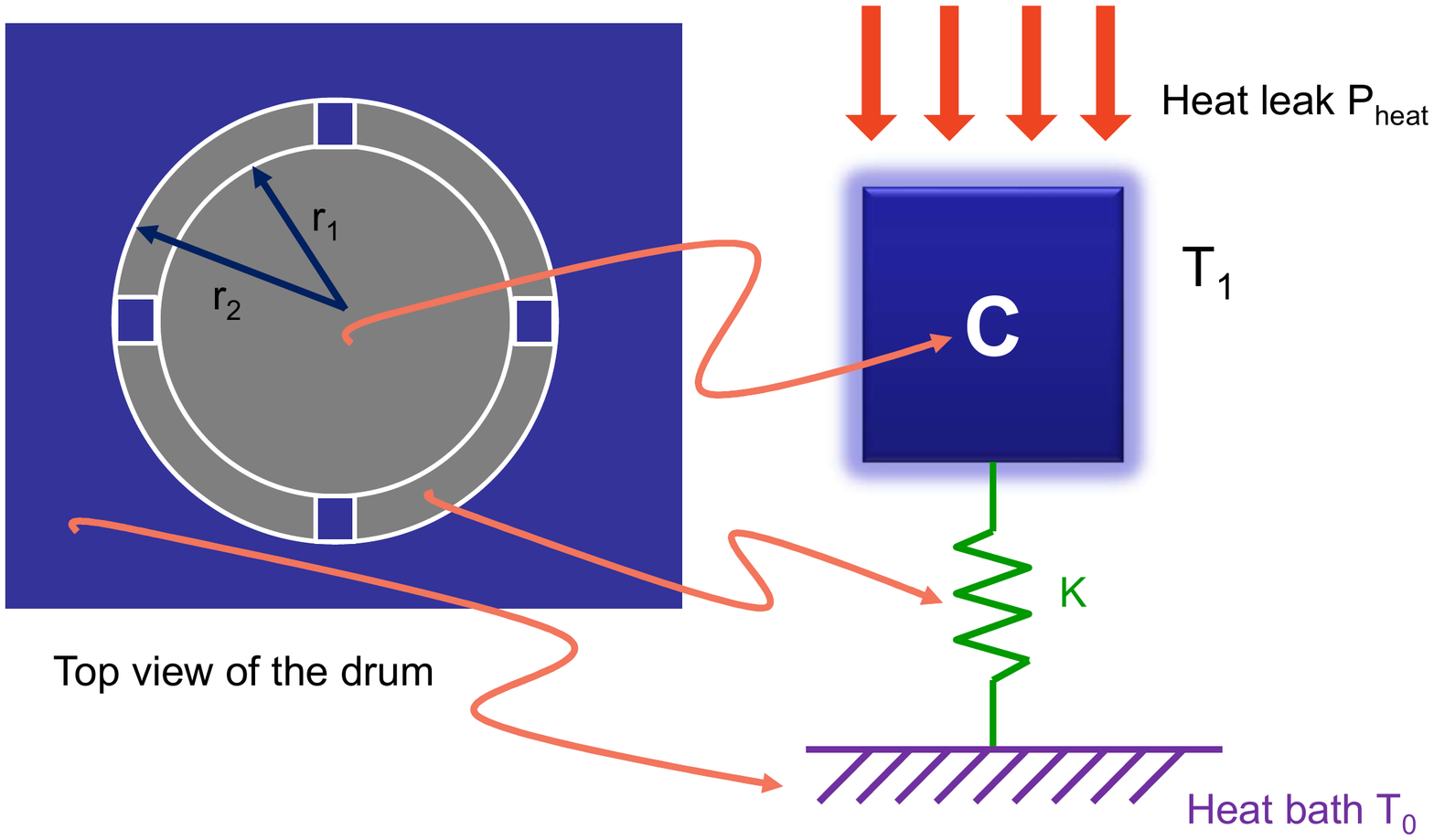}
		\vspace*{-2.5cm}
	\end{center}
	\caption{{\bf Thermal model.} Left: schematics (not to scale) of the suspended aluminium disk (at temperature $T_1$) and the (cut) torus connecting the disk to the heat bath (the sapphire substrate, at $T_0$), see image in Fig. \ref{fig_1}. Right: thermal model of the suspended disk membrane of heat capacity $C \approx C_{membrane}$ and thermal conductance $K \approx K_{torus}$ of the torus (we neglect then the small cuts). ``heat leak'' stands for {\it all} the heat sources acting on the membrane (see discussion in text). }
	\label{fig2}
\end{figure}

The second important calculation is about estimating the thermal conductance of the structure in order to evaluate its thermalisation time (and potential thermal gradients).
The simplified model we consider is shown in Fig. \ref{fig2}. 

\begin{figure}
	\begin{center}
		\includegraphics[width=11cm]{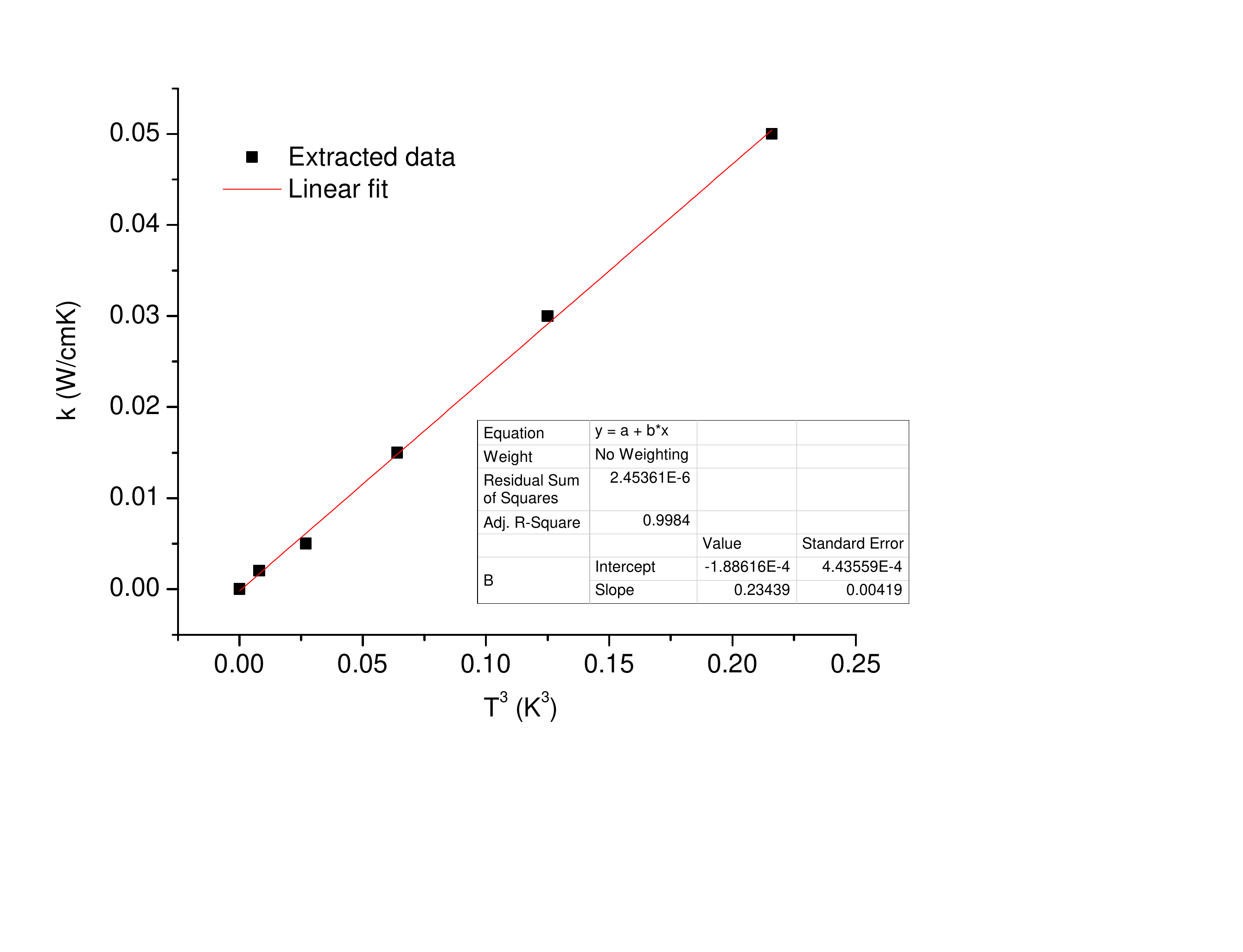}
		\vspace*{-2.5cm}
	\end{center}
	\caption{{\bf Thermal conductivity fit.} Bulk thermal conductivity of aluminium as a function of $T^3$ extrapolated from Zavaritskii \cite{zav} up to high temperatures enabling to calculate the power law $k\approx$ 23.4~$T^3$ W.m$^{-1}$.K$^{-1}$ (Tab. \ref{table1}).}
	\label{fig3}
\end{figure}

The phonon thermal conductance is related to specific heat and phonon mean-free-path through the kinetic equation (coming from the Boltzmann transport equation at low temperature, see e.g. Ref. \cite{Cleland}):
\begin{equation}
k= \frac{1}{3}  c_p \Lambda v_s .
\label{kinetic}
\end{equation}
From this equation, we can calculate the bulk phonon mean-free-path $\Lambda$ if we know the bulk (experimental) thermal conductivity $k$ and the specific heat $c_p$. The measured bulk thermal conductivity is taken from Zavaritskii \cite{zav} (Fig. \ref{fig1}) by fitting the lower part of the curve below 0.2~K (see Tab. \ref{table1}; and the power law extrapolation in Fig.~\ref{fig3}).
This gives a mean-free-path for bulk phonons of $\Lambda=2.5 \times 10^{-2}$~m (2.5~cm, and independent of temperature since both $k$ and $c_p$ have the same $T^3$ dependence).
From this result, we shall simplify the problem by considering the membrane as isothermal at temperature $T_{1}$, and assume that the temperature gradient (if any) happens at the torus region. We should now define the thermal conductance of this zone.

When we build a system where one dimension is smaller than this mean-free-path, the phonons will scatter on the rough surfaces that confine them; then $\Lambda$ will be reduced. 
The new mean-free-path can be obtained following the Casimir model \cite{Ziman,Casimir1938,Heron}, a rather simple model, where we assume that the phonons are fully scattered on surfaces (a sort of ``phonon black body''). 
This is indeed what happens in the torus region, where the thickness $e_p$ is only about 100$~$nm.
Then the new (``confined'') $\Lambda_{torus}$ is essentially given by the smallest dimension of the thermal conduction channel (a ``worse estimate'', here the thickness already quoted, much smaller than the radii $r_1$ and $r_2$).
Taking as an order-of-magnitude approximation $\Lambda_{torus} \approx e_p$ (then $10^4$ times smaller than the bulk one), the ``confined'' thermal conductivity writes:
\begin{equation}
k_{Al\,nano}= \frac{1}{3}  c_p \Lambda_{torus} v_s ,
\label{kinetic2}
\end{equation}
where $c_p$ is the bulk specific heat in J.m$^{-3}$.K$^{-1}$, and $\Lambda_{torus}$ the new phonon mean-free-path in the {\it Al} torus. This is giving a thermal conductivity of $k_{Al\,nano}= 10^{-4}\, T^3$~W.m$^{-1}$.K$^{-1}$.

\subsection{Characteristic thermalisation time}

The characteristic thermalisation time $\tau_{th}$ is given by $\tau_{th}=C_{membrane}/K_{torus}$.
We define the membrane heat capacity:
\begin{equation}
C_{membrane}= c_p V ,
\label{cmemb}
\end{equation}
with $V=e_p \pi r_1^2$ the volume of the suspended part in Fig. \ref{fig2} (with $r_1 = 7~\mu$m).
The numerical value is then $C_{membrane}=6 \times 10^{-18} \, T^3$~J.K$^{-1}$.
In order to estimate the thermal conductance $K_{torus}$ of the {\it Al} membrane to the heat bath, we make a simple ``worse estimate'' calculation assuming that the thermal gradient develops on the full width of the torus, from $r_1$ to $r_2$ (see Fig.~\ref{fig2}).
Since $r_1, r_2 \gg \Lambda_{torus}$, we shall assume that the conduction is {\it diffusive} 
in the torus.
Noting that the dominant phonon wavelength $\lambda_{dom.} \approx 2.23 \, \hbar v_s /(k_B T)$ is of the order of $e_p$ around 1$~$K, we can treat the problem as  pure 2D  transport, leading to the simple result \cite{Maasilta}:
\begin{equation}
K_{torus}= \frac{2 \pi e_p}{ln(r_2/r_1)} k_{Al\,nano} ,
\label{kalnano}
\end{equation}
with the two radii $r_1=$7~$\mu$m and $r_2=$10~$\mu$m. 
We thus get as an estimate $K_{torus}=1.6 \times 10^{-10} \, T^3~$W.K$^{-1}$.

We obtain from our estimates $\tau_{th}= 4\times 10^{-8}~$s (that is 40 nanoseconds), independent of temperature.
This is obviously a rough estimate, giving however the correct order of magnitude. 
It tells us that the bath itself can accept/give thermal energy at {\it  a very fast rate}, up to around a GHz.

\subsection{Kapitza thermal resistance and heat dissipation within the aluminium}

The simple model assumes that the sapphire substrate and the copper cell (which are macroscopic) are well anchored at the cryostat temperature $T_0$. 
But there is an extra thermal boundary resistance that we did not include yet: the Kapitza resistance between the aluminum torus and the sapphire, which is due to the {\it acoustic mismatch} between the two solids \cite{PobellBook}.
We take from Swartz \textit{et al.} \cite{swartz} the thermal conductance per surface given by: $k_{ac.\,mis.}=0.1 \,T^3~$W.cm$^{-2}$.K$^{-1}$, then $K_{Kapitza}=k_{ac.\,mis.} \times S_{torus}$ with $S_{torus}$ the area of the torus. From our dimensions, we calculate a contact surface $S_{torus} = 1.6 \times 10^{-6}~$ cm$^{2}$, which leads to a Kapitza thermal conductance of $1.6 \times 10^{-7}\, T^3$~W.K$^{-1}$. 
This value is much larger than $K_{torus}$, which means that we can indeed safely neglect it in our discussion.

At the lowest temperatures, a thermal gradient might develop between the bulk (at temperature $T_0$) and the aluminum drum-head (at $T_1$):
\begin{equation}
T_{1}-T_{0}=P_{heat}/K_{torus} ,
\label{gradient}
\end{equation}
with $P_{heat}$ the {\it total} power that is absorbed by the suspended membrane (Fig. \ref{fig2}). 
The question that naturally arises is: what mechanisms eventually limit $P_{heat}$? 
Talking only about known (and somehow steady, thus not considering the ``spike'' issue here) mechanisms, one immediately thinks of the microwave power injected, or the imperfect r.f. filtering that reaches the sample. The former is measured to be negligible (see Section \ref{microwave}), while the latter couples essentially to the electrons (see following subsection), not to the phonons. Black-body radiation is also negligible, since the whole chip is enclosed in a copper cell thermalised to the nuclear stage, which acts as a shield.
Heating due to cosmic rays or natural radioactivity (from the building itself) is also an eligible mechanism \cite{ParpiaRad}, although almost impossible to estimate. Since the cross section of the device (and its thickness) is extremely small, we shall also neglect it.

Beyond {\it radiation}, one can also think about adsorption of gas particles: using $^4$He as exchange gas in the 4$~$K pre-cooling process, if the remaining vacuum in the can is not low enough at cryogenic temperatures, atoms may desorb from hot parts and condense on the cold ones (i.e. the chip), creating thus an almost permanent heat leak \cite{PobellBook}. This phenomenon is well known, and has been taken care of.
Finally, the last external mechanism that people take seriously in consideration is {\it mechanical vibrations}. This is known to be detrimental to low frequency devices, generating huge heat loads if not properly handled by complex suspension systems \cite{oosterkamp}.
For high frequency devices, this seems to be much less of a problem: they do thermalise to tens of milliKevin with conventional cryogenics \cite{Teufel}. In our case, the whole cryostat is suspended on air-mounts with about a ton of concrete anchoring, which is required for cryogenic purposes (the copper stage shall not vibrate in the 7$~$T field, otherwise eddy currents would create a huge heat leak) \cite{PobellBook}.

We are therefore left with {\it internal} sources of heat within the {\it Al} membrane, which can become (very) relevant at ultra-low temperatures \cite{PobellBook}.
These are time-dependent (but with rather long time-scales, so that they always contribute to some extent), and originate in H$_2$ inclusions, structural defects (e.g. grain boundaries in the metal layer, or TLSs in glasses) that can relax, or even radioactive impurities (citing documented effects). 
It is impossible to estimate quantitatively this contribution, apart from fitting results from an actual experiment. 
We shall thus simply consider a reasonable upper bound that supports our finding: about 0.1$~$nW/kg (after a few weeks of experiments below 1$~$K) \cite{PobellBook}.
This number corresponds to a negligible heating of the mechanical element above 1$~$mK, and about 100$~\mu$K temperature increase only around 600$~\mu$K.

\subsection{Conduction electrons}

We shall conclude our discussion on thermal properties by a quick look at the electron bath in the (superconducting) aluminium.
The electron bath decouples from the phonon bath in the membrane following the equation \cite{Roukes}:
\begin{equation}
P_{e-}= V g_{e-ph} \left( T^{5}_{e-}- T^{5}_{ph} \right) ,
\label{T5}
\end{equation}
where $P_{e-}$ is the power arriving on the electrons, $T_{e-}$ and $T_{ph}$ are respectively the electronic and phononic temperatures, and $V$ the volume of the drum. $g_{e-ph}$ is the electron-phonon coupling constant, taken here to be $0.4 \times 10^9~$W.K$^{-5}$m$^{-3}$ \cite{Wellstood}.
From the discussion of the previous section, we can take $T_{ph} = T_1 \approx T_0$. 

Since these electrons hold a vanishingly small specific heat \cite{BCS}, any small source of heat $P_{e-}$ easily keeps them hot: this is a major problem in quantum electronics, which required a lot of work over the years to prevent superconducting circuits to be poisoned by quasi-particles (i.e. non-paired electrons)
\cite{Jukka_NO_QP}.
As an illustration here, a rather correct filtering of r.f. noise leading to a heat leak of 1$~$fW would create a base temperature for the electrons of 40$~$mK, while phonons {\it do cool} below 1$~$mK. An extremely good filtering producing $P_{e-} \approx 1~$aW would still keep $T_{e-}$ at about 10$~$mK, among the best electronic temperatures reported in nanostructures.

\appendix
\section*{DATA AVAILABILITY}
The data that support the findings of this study are openly available at the following URL/DOI: 
\small{\underline{https://cloud.neel.cnrs.fr/index.php/s/CnnYPKn8XHYZgXa}}.

\end{document}